석 사 학 위 논 문

Master's Thesis

# 다중수단 연계를 고려한 공유교통체계의 매칭 문제

Multi-modal Matching Problem of Shared Mobility

2016

우 수 민 (禹 秀 旻 Woo, Soomin)

한 국 과 학 기 술 원

Korea Advanced Institute of Science and Technology

석 사 학 위 논 문

# 다중수단 연계를 고려한 공유교통체계의 매칭 문제

2016

우 수 민

한 국 과 학 기 술 원

건 설 및 환 경 공 학 과

# 다중수단 연계를 고려한 공유교통체계의 매칭 문제

우 수 민

위 논문은 한국과학기술원 석사학위논문으로
학위논문 심사위원회의 심사를 통과하였음

2016년 6월 10일

심사위원장　여 화 수　( 인 )

심 사 위 원　김 영 철　( 인 )

심 사 위 원　장 기 태　( 인 )

# Multi-modal Matching Problem of Shared Mobility

Soomin Woo

Advisor: Hwasoo Yeo

A thesis submitted to the faculty of
Korea Advanced Institute of Science and Technology in
partial fulfillment of the requirements for the degree of
Master of Science in Civil and Environmental Engineering

Daejeon, Korea
June 10th, 2016

Approved by

___________________________
Hwasoo Yeo
Associate Professor of Civil and Environmental Engineering

The study was conducted in accordance with Code of Research Ethics[1].

___________________________
1) Declaration of Ethical Conduct in Research: I, as a graduate student of Korea Advanced Institute of Science and Technology, hereby declare that I have not committed any act that may damage the credibility of my research. This includes, but is not limited to, falsification, thesis written by someone else, distortion of research findings, and plagiarism. I confirm that my dissertation contains honest conclusions based on my own careful research under the guidance of my advisor.

초 록

기존의 승차공유 시스템은 승용차 등의 단일 수단 기반으로만 제안되고 있어 차량 이동에 대한 수요와 공급이 비대칭일 때 적용이 제한적이다. 이에 본 논문에서는 다중수단을 기반으로 하는 승차공유의 적용을 위하여 대중교통과 승용차 간의 환승을 허용하는 운전자와 승객 간의 매칭 프레임워크를 개발하였다. 이전의 논문과는 다르게 대중교통과 스케줄의 유연성을 함께 적용함으로써 주어진 사용자 환경에서 프레임워크의 매칭율이 증가하였다. 특히, 승차공유 시스템에서 운전자가 이동 수요자보다 상당히 적을 때 대중교통에 대한 환승을 허용하고 스케줄의 유연성을 조정함으로써 매칭율이 높은 수치에서 빠르게 수렴하게 하였고, 이는 승차 공유 시스템의 실제 적용 가능성을 높일 수 있다.

핵 심 낱 말   공유교통체계, 다중수단, 승차공유, 유전자알고리즘, 매칭알고리즘

Abstract

Rideshare is one way to share mobility in transportation without increasing traffic demand, where travel mobility and usage of vehicle capacity can be improved. However, current literature on rideshare has allowed only one-modal trips and may be limited in the matching efficiency, especially when there is a large gap between the supply and demand of mobility. Therefore, the objectives of this paper are first to develop a multi-modal matching framework of shared mobility with public transportation to maximize the performance of a rideshare system, and second to evaluate the effect of the public transportation and of the schedule flexibility on the matching efficiency. To fulfill the first objective, a multi-modal matching framework is developed to allow rideshare with both private and public vehicles with detailed design of detour, using Genetic Algorithm. Also for the second objective, the effects of public transportation and schedule flexibility are evaluated with a simplified network of Sioux Falls. The results show that public transportation helps the match rate slightly at a low supply of private vehicle, but this must be evaluated for practical implementation as different cities may bring different results. Also, a larger schedule flexibility helps greatly in increasing match rate even at a lower supply level. As well, the planning subject of time schedule is benefited more with larger schedule flexibility, in this paper the drivers, on the matching efficiency. Moreover, a rideshare system with private vehicles outperforms a public transportation system, possibly due to the rigid route of public transportation that takes no detour burden. This confirms the need for a flexible design of sharing mobility, as can be fulfilled with the multi-modal matching framework developed in this research.

Keywords   Shared mobility, multi-modal transportation, rideshare, Genetic Algorithm, matching algorithm

# Table of Contents

# List of Figures

# List of Tables

# Chapter 1. Introduction

## 1.1 Rideshare System

Rideshare is a joined trip of drivers and riders, whose requests are matched to meet every participant's need of travel. This system may help alleviate one of the main urban problems, traffic congestion, by increasing efficiency of using vehicle capacity with more riders in each car, therefore decreasing number of vehicles and traffic density on the road[1]–[33]. Another possible advantage is the increased mobility of travelers without vehicles, possibly with less cost than taxi and more flexibly than public transportation. With a wide spread of mobile phones equipped with GPS, the rideshare system has a better advantage by using accurate location information of participants in real time and avoiding extra installation of infrastructure, such as meeting points.

Note that rideshare is different from other vehicle-to-rider matching systems, such as taxi-sharing, car-pooling and shared autonomous vehicles. One distinct property of rideshare is that in ridesharing, the drivers have their own trip to pursue including origin, destination, and time schedule. This creates a large difference of rideshare from taxi-sharing and shared autonomous vehicles, where drivers are not confined to the spatial and temporal limitations of its own so that matching of riders and vehicles is a problem much more relaxed[23], [34]–[36]. Additionally, car-pool is similar to rideshare but the term 'car-pool' is used often for repetitive sharing with longer time to plan for example in terms of days, such as commuting. The term 'rideshare' is used more often for dynamic sharing with shorter time to plan in terms of minutes or hours[22].

Even though it has a potential to improve the mobility of a city, a rideshare system is faced with various challenges, such as pricing issues, incentive issues, and safety issues. One of them is the 'chicken-and-egg problem', where the participation rates of drivers and riders are co-dependent in terms of the match rate [16], [22], [23], [28]. When the size of drivers and the size of passengers are unequal, the group in larger size suffers from a low success rate in finding a match and may be discouraged to join the system again. However, the larger group is a necessary element in increasing the participation of the smaller group by providing a successful match rate. Only when the smaller group increases in size, the matching efficiency for the larger group can increase. This intermingled relationship on match rate may hinder the system from providing a constant and successful match rate. Therefore, it is necessary to maximize the match rate at a given pool of participants even when it is unbalanced.

## 1.2 Rideshare Matching Problem

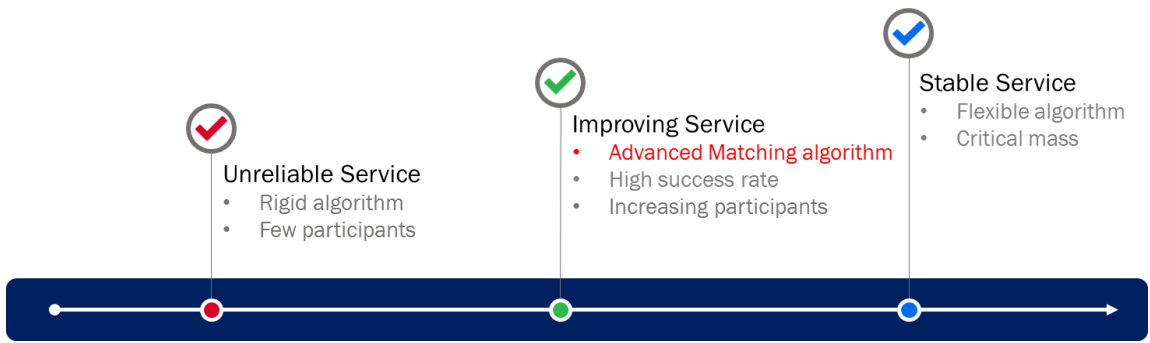

**Figure 1 Rideshare Implementation**

One approach to maximize the match rate and to actively invite participants to the system is to advance the matching algorithm. The rideshare matching problem is a problem of matching drivers and riders for a joined trip, which must satisfy their spatial and temporal requests like origin, destination, and travel schedule. Refer to Figure 1 for explanation of how it may help implementing the rideshare system in practice. If there is a small group of participants and a rigid algorithm to create matches, the rideshare matching service can only provide unreliable service with a low success rate. However, as the matching algorithm is improved, a higher success rate can be provided even at a poor participation rate with imbalance. Then more participants may join in with a better matching probability. In the end with the flexible algorithm, the service may reach stability with a constant, high success rate at critical mass of participants.

So far the researchers on rideshare have attempted to give more flexibility to the rideshare rule and increase the solution space for matching algorithm in order to increase matching efficiency at a given set of requests [1]–[5], [7], [8], [10]–[21], [23]–[27] . The relaxation of rideshare rule can be categorized into four features – matching configuration, detour burden, vehicle pool augmentation, and schedule flexibility. Matching configuration considers how many drivers or passengers a participant can be matched to, detour burden considers who takes the burden of detour incurred in a joined trip and how, vehicle pool augmentation considers the choice spectrum of ride offers like private vehicle or non-private vehicles, and schedule flexibility considers the willingness of participants to travel longer for their trips. The first three features are closely related to the matching algorithm design and the last is related to the parameter of the algorithm.

However, the current literature has only developed a matching model that allows one-modal rideshare trips. Some studies that allow transfers in a journey or multiple vehicle types still have only allow a single mode for travel. The opportunity here is to develop a matching framework that fully utilizes another transportation mode, such as public transportation, which is already installed in most cities with high reliability. Then, this integrated framework has a larger potential than the current literature because it opens up the dimension of shared mobility with multiple mode choices from conventional rideshare. After the framework is developed, this study may evaluate the potential of this shared-mobility system further by its matching efficiency at various supply levels. Therefore, the objective of this research is given as follows.

1.3  Objective Statement

The first objective is to develop a multi-modal rideshare matching framework with public transportation. This is an expansion from a conventional rideshare to a shared-mobility with multiple mode choices, by relaxing match configuration, detour burden, and vehicle pool augmentation of rideshare rule. Also, this framework should achieve a successful performance with a high match rate in a reasonable execution time.

The second objective is to evaluate the performance of the rideshare matching framework for the effect of vehicle pool augmentation as well as the effect of schedule flexibility of participants. The objective function of successful match rate is studied to provide important insights on the implementation and operation of rideshare system.

1.4  Paper Structure

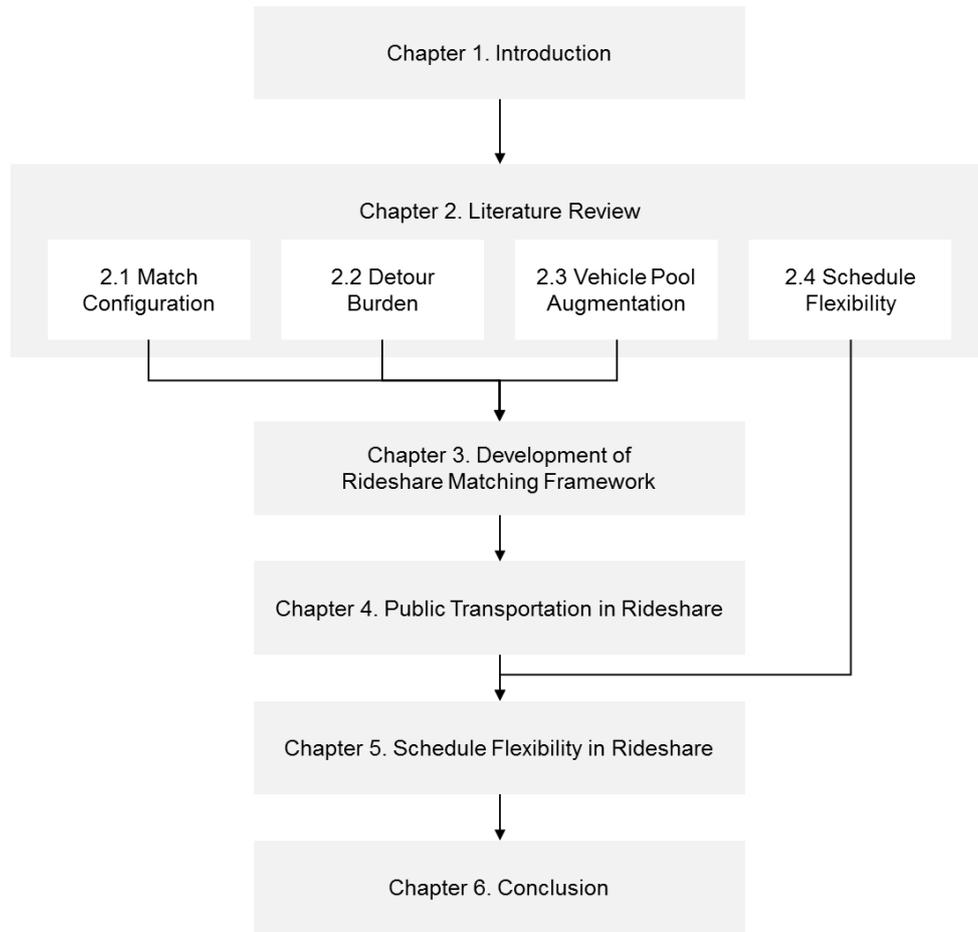

**Figure 2 Paper Structure**

This research is structured as follows as shown in Figure 2. Chapter 2 has literature review on the rideshare matching algorithms on the categories of match configuration, detour burden, vehicle pool augmentation, and schedule flexibility. The first three features will be used to develop a rideshare rule and matching framework in Chapter 3. Chapter 4 evaluates the performance of the proposed framework with the effect of augmented vehicle pool, whereas Chapter 5 evaluates the performance with the effect of schedule flexibility of participants. Both chapters attempt to provide guidelines for a rideshare system in practice. Finally, Chapter 6 highlights the findings of previous chapters and suggests future work needed after this study.

# Chapter 2. Literature Review

With wide spread of mobile phones equipped with GPS, the rideshare system has a better potential of dynamic matching with much more accurate and flexible location. One of the challenges faced by rideshare is the 'chicken-and-egg' problem, where the participation rate of drivers and passengers depend on, but also affected by, each other with the match rate [16], [22], [23], [28]. In order to maximize the matching efficiency, many papers on rideshare have attempted to improve the matching algorithm with unequal pool of drivers and passengers. In this chapter, the current literature on rideshare matching algorithm is assessed for its shortcomings and the opportunity for research.

The author has studied and categorized the literature into four different features to relax the spatial and temporal constraints of rideshare matching, thereby increasing the solution space and matching efficiency. The four relaxation features are match configuration, detour burden, vehicle pool augmentation, and schedule flexibility. The first three are often explored and studied to build a rideshare matching framework, whereas the last has not been actively studied in the influence on the algorithm design. Therefore, the next four sections explain in detail these features with the current literature. This chapter is finished with the shortcomings of the literature and an opportunity for improving rideshare matching algorithm. The literature investigated by the author in this chapter is summarized in Table 1.

**Table 1 Current Literature in Rideshare Matching Algorithm**

| Paper | Year | Matching Configuration | | Detour Burden | Vehicle Pool Augmentation |
|---|---|---|---|---|---|
| | | Number of passengers per driver | Number of drivers per Passenger | | |
| [1] | 2009 | Multiple | One | Drivers by detour | - |
| [2] | 2009 | One | Multiple | Mutual, drivers by detour and passengers by walking to transfer location | - |
| [3] | 2011 | One | One | Driver by detour | Shifter (Request as both passenger and driver) |
| [4] | 2011 | One | One | Drivers by detour | - |
| [5] | 2012 | Multiple | One | Drivers by detour | - |
| [8] | 2012 | One | Multiple | Drivers by detour | - |
| [7] | 2012 | Multiple | Multiple | Drivers by detour | - |
| [11] | 2013 | One | Multiple | Drivers by detour | - |
| [12] | 2013 | Multiple | One | Drivers by detour | - |
| [13] | 2013 | Multiple | One | Drivers by detour | Taxi |
| [15] | 2014 | Multiple | One | Drivers by detour | Shifter (Request as both passenger and driver) |
| [16] | 2014 | Multiple | One | Mutual by hubs | - |
| [17] | 2014 | Multiple | One | Drivers by detour | Taxi |
| [10] | 2014 | Multiple | One | Drivers by detour | - |
| [19] | 2015 | One | One | Mutual by intermediate locations | - |
| [18] | 2015 | One | One | Mutual by relay stations | - |
| [20] | 2015 | Multiple | One | Drivers by detour | - |
| [21] | 2015 | One | One | Drivers by detour | Dedicated drivers |
| [23] | 2015 | Multiple | Multiple | Drivers by detour | - |
| [14] | 2015 | Multiple | One | Drivers by detour | Taxi |
| [24] | 2015 | Multiple | Multiple | Passengers by multi-hop | - |
| [25] | 2015 | Multiple | One | Drivers by detour | - |
| [26] | 2016 | Multiple | One | Mutual, both by detour | - |
| [27] | 2016 | Multiple | Multiple | Mutual, both by hub | - |

2.1 Match Configuration

This paper defines match configuration as the format of how drivers and passengers incorporate each other in a joined trip. This category describes how many passengers a driver can deliver and how many drivers a passenger can take. Match configuration is described on both aspects of drivers and passengers.

On the driver's aspect, it is possible for a driver take a single passenger request or multiple as shown in Table 2. If he takes a single request, the joined trip starts from the driver's origin, passenger's origin and destination, ending with the driver's destination [3], [4], [18], [19], [21]. The single request can be a passenger or a group of passengers requesting as a team. In case of the latter, the match success also depends on the seat capacity of driver but the required visits by the driver is the same as the former. The latest papers usually take the multiple request configuration, where the joined trip may have separate travels or combined travels as shown in the figure [1], [5], [7], [10], [12]–[17], [20], [23]–[27]. In case of separate travels, the vehicle contains passengers of each request, in other words, a drop-off of one request must be finished before the pickup of another request. In case of combined travels, the vehicle can contain passengers of multiple requests, in other words, a request may not be fulfilled before another request starts as long as one request's pickup is before its drop-off. The multiple requests and the combined travels are more flexible with larger solution space of matching than the single requests and separate travels, respectively.

**Table 2 Match Configuration (Driver's Aspect)**

| Number of Requests | Number of Passengers | Travel Type | Example of Driver's Visits |
|---|---|---|---|
| **One** | One | Separate | Driver's origin → Passenger pickup → Passenger drop-off → Driver's destination |
| | Multiple | Separate | Driver's origin → Passenger pickup → Passenger drop-off → Driver's destination |
| **Multiple** | Multiple | Separate | Driver's origin → Passenger A pickup → Passenger A drop-off → Passenger A pickup → Passenger A drop-off → Driver's destination |
| | Multiple | Combined | Driver's origin → Passenger A pickup → Passenger A pickup → Passenger A drop-off → Passenger A drop-off → Driver's destination |

On the passenger's aspect, a passenger can take a single driver or multiple drivers to finish the trip as shown in Table 3. The former is called 'single-hop' and the latter, 'multi-hop'. Many papers have used the single-hop configuration because it has much lower complexity in matching constraints and route design [1], [3]–[5], [10], [12]–[21], [25], [26]. However, the multi-hop trips have a much larger solution space and flexibility as the other papers argue, since it has a better chance of finding a ride where the passenger wants to go [2], [6]–[8], [11], [23], [24], [27]. Therefore, multi-hop is a spatial relaxation of rideshare constraint. A multi-hop configuration combines routes of multiple driver offers, which are confined by their own limits of detour. Also by transferring, multi-hop open up the possibility of multi-modal trips, for instance mixing private and public vehicles to fulfill the passenger's journey. If the algorithm finds matches that meet the detour limit and the time schedule of passengers and drivers, multi-hop can be more beneficial than single-hop for successful match rate.

**Table 3 Match Configuration (Passenger's Aspect)**

| Number of Drivers | Example of Passenger's Visits |
|---|---|
| **One ('single-hop')** | Passenger's origin → Driver pickup → Driver drop-off → Passenger's destination |
| **Multiple ('multi-hop')** | Passenger's origin → Driver A pickup → Driver A drop-off → Transfer → Driver B pickup → Driver B drop-off → Passenger's destination |

The match configuration of n-passengers to m-drivers (where n,m≥1) is argued to be slow for practical applications but with potential for matching efficiency [23]. However, a meta-heuristic method like Genetic Algorithm may be used to execute rideshare matching in this configuration and to maximize matching probability with given request set in a reasonable time frame.

## 2.2 Detour Burden

This paper defines detour burden as the additional travel distance incurred to join trips of participants, whose shortest travel paths may not coincide. The detour burden for a driver is calculated from the deviation of his original path to finish the joined trip, whereas for a passenger it is calculated from the amount of travel distance on its own. This category considers who takes the detour burden, i.e. driver or passenger, and how the burden will be taken, i.e. full responsibility or partitioned.

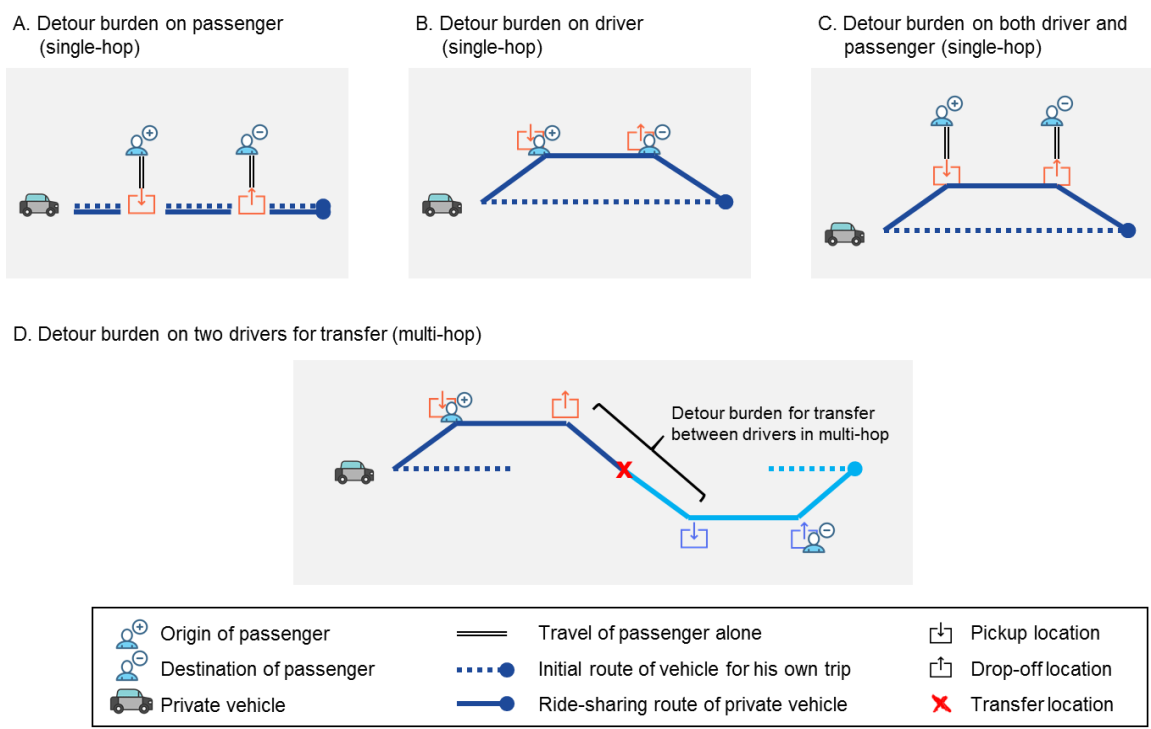

**Figure 3 Different Carriers of Detour Burden**

First consider only the case of single hop, where detour burden is either on passenger or driver. The detour burden can be taken by passengers, by drivers, or mutually by both passengers and drivers as shown in Figure 3-A to Figure 3-C. When the passenger takes full burden of detour, he must arrive at the pickup point that is on the driver's initial route for the driver's own journey as shown in Figure 3-A.. This type is only seldom used as the passenger without vehicle has to complete his journey with large penalty [24]. Especially if the rideshare requests are distributed sparsely, the penalty on the passenger may be too large and discourage them from forming a match. This type must consider the passenger's constraint for this extra effort, for example the maximum distance he is willing to walk for the burden.

When the driver takes full burden of detour, which is often the case for most of literature, the pickup and drop-off of passengers must coincide with the origin and destination of passengers, respectively [1], [3]–[5], [7], [8], [10]–[15], [17], [20], [21], [23], [25]. Shown in Figure 3-B, this type is where many papers insist on the advantage of rideshare with the door-to-door feature, especially for residential area far from public transportation access. Also, the driver's constraint for this extra effort to detour should also be considered.

When the detour burden is taken mutually by both drivers and passengers, intermediate location is needed like Figure 3-C. The pickup and drop-off locations may not be the origin and destination of passengers; however, they also may not be on the driver's initial route for his own journey. Fixed meeting locations are used in some studies that concentrate the dispersed locations of rideshare demands and help matching efficiency [16], [18], [19], [27]. However, flexible meeting location is also possible [2], [26]. As well, constraints of both drivers and passengers for this extra effort to detour must be checked in creating a match.

The above cases in Figure 3-A to C of detour burden only consider the origin and destination of the passengers in single-hop, i.e. burden division between driver and passenger. For multi-hop trips, however, a detour burden also arises from the transfer process between the drivers assigned to a passenger. As shown in Figure 3-D, there may not be an intersection of both drivers' routes convenient for transfer and the two drivers must decide on where to meet up. To the best of author's knowledge, no paper has considered the detour burden of transfer in a multi-hop trip. However, this detour must be considered in a design of multi-hop rideshare. This negotiation of detour burden between the drivers may be rule-based, for instance full responsibility by the dropping-off driver, or merit-based, for instance less detour for the driver with more passengers. Again, the constraints of drivers on detour must be checked.

To summarize, for the origin and destination of passengers, detour burden on private drivers rather than on passengers seem to have been supported by most of the literature, shown in Figure 3-B. For the transfer process of multi-hop configuration, a study on the detour burden between drivers is lacking and should be evaluated for the details of practical application of rideshare.

2.3 Vehicle Pool Augmentation

This paper defines vehicle pool augmentation as an increased vehicle pool for service. Conventionally, a rideshare system would register only the private drivers as viable candidates. However, some papers have developed a matching methodology to increase the use of a given vehicle pool. One example is integration of taxi-sharing and rideshare, where passengers can be assigned to either a private vehicle for a joined trip or other passengers to share a taxi [13], [14], [17]. Another example is a flexible role of driver and rider, called 'shifters', where a request is sent to the matching algorithm as both potential driver and rider[3], [15]. This unifies the request pool of driver and riders, reducing the size gap between drivers and riders. Also, the possibility of using dedicated drivers has been studied for boosting rideshare implementation, as a ground supply of rides [21].

Public transportation is one of the possibilities on increasing the vehicle pool in rideshare as mentioned in a few papers [3], [16], [29]. It has advantages with a high reliability of service with predetermined schedule at a low cost. It is installed already in most cities and despite its inefficiency, it can help fill the spaces where driver offers are sparse. No paper to the best of author's knowledge has developed a rideshare matching algorithm that integrates public transportation as a vehicle option. If public transportation is integrated with rideshare, multi-modal trips can be designed. Passengers can use the public transportation to travel from origin and destination where offers are scarce, with small cost of inefficiency.

2.4 Schedule Flexibility

Schedule flexibility in this paper is the participants' willingness to travel longer than their direct travel time, as defined in a rideshare matching study by Agatz et al [3]. This temporal relaxation of participants' on rideshare matching has been studied to show its potential on increasing the efficiency of rideshare matching [3]. Also it has been used as a parameter for the matching algorithm [27]. However, it has not been evaluated on how it affects the matching efficiency for a complex matching configuration like such as multi-hop, and how the rideshare matching can be designed to maximize the benefit of schedule flexibility. Especially at an unequal supply and demand, this parameter may help increase the matching efficiency further.

2.5 Opportunity

The current research has studied the rideshare matching algorithm on each of four relaxation features, but there is still a room for improvement. It is possible to develop a rideshare matching framework of most advanced match configuration, together with additional vehicle pool of public transportation. The matching framework will permit multi-modal trips and expand from the conventional rideshare system to a larger scope of sharing mobility. This framework can be designed by relaxation of match configuration, detour burden, and vehicle pool choice. Then, it is possible to evaluate the effect of public transportation and the behavior of participants on rideshare on matching performance.

Therefore, this research attempts to develop a rideshare matching framework that use a) a match configuration of multiple passenger and multiple drivers, b) detour burden consideration on origin, destination, and transfer process, and c) vehicle pool augmentation with public transportation. Additionally, the effect of public transportation and the schedule flexibility among participants can be evaluated to give guidelines for the implementation of rideshare. In the following chapter, the rideshare matching framework is developed with the fully relaxed rideshare rule.

# Chapter 3. The Framework for Multi-modal Rideshare Matching

In this chapter, a framework for multi-modal rideshare matching with private and public vehicles is developed. The first section describes the rideshare rule for multiple-passenger, multiple-driver configuration with public transportation, where the relationship between three relaxation features – match configuration, detour burden, and vehicle pool augmentation – are understood in terms of the rideshare rules and adopted for the framework. Next, a formal definition of the rideshare matching problem is given. Then, the rideshare matching framework is developed in detail using Genetic Algorithm and its performance is verified. This chapter is finished with a conclusion for the development of the framework.

## 3.1 The Rideshare Rules

The rideshare in this paper follows rules as given below. The rules are first described generally, then in more detail with match configuration, vehicle pool augmentation, and detour burden, regarding the relationship between the three features.

First, the general rule for rideshare is described. The rideshare requests from drivers and riders are given and the matching framework uses the known requests in the system. The term 'participant' in this paper only describes private vehicle drivers and passengers, but not public vehicles. Request information of a private driver is to entail their origin, destination, earliest departure time, latest arrival time, and the seat capacity. Request information of a rider is to entail their origin, destination, earliest departure time, latest arrival time, and the limit of walking range in case of taking a public transportation in the trip. Public transportation can be considered available for the multi-modal trips of rideshare, however is not included in the calculation of ride supply as it is assumed as a given parameter in a system. The information of public transportation required for matching is the fixed set of visit stations, fixed period of dispatch time, and the first and last dispatch time.

Another different rule of this framework to the other rideshare studies is that it is much more flexible in terms of the intentions of participants. Note that this matching framework assumes that the participants take the rideshare match as suggested, if their spatial and temporal constraints are met, i.e. origin, destination, and the time schedule, as well as other individual allowances, i.e. the seat capacity of driver or the limit of walking range for passengers. This is different from many other matching algorithms that specifically give inefficiency of joined trips a large penalty, for example for a transfer or detour burden, where rideshare matching is driven by the sole motive of finishing their joined trips in the most efficient way possible. In comparison, this framework considers that the participants are more accepting of the inefficiency of joined trips by assuming their selfish intentions and select their spatial and temporal constraints as they can maximally allow rather than they need.

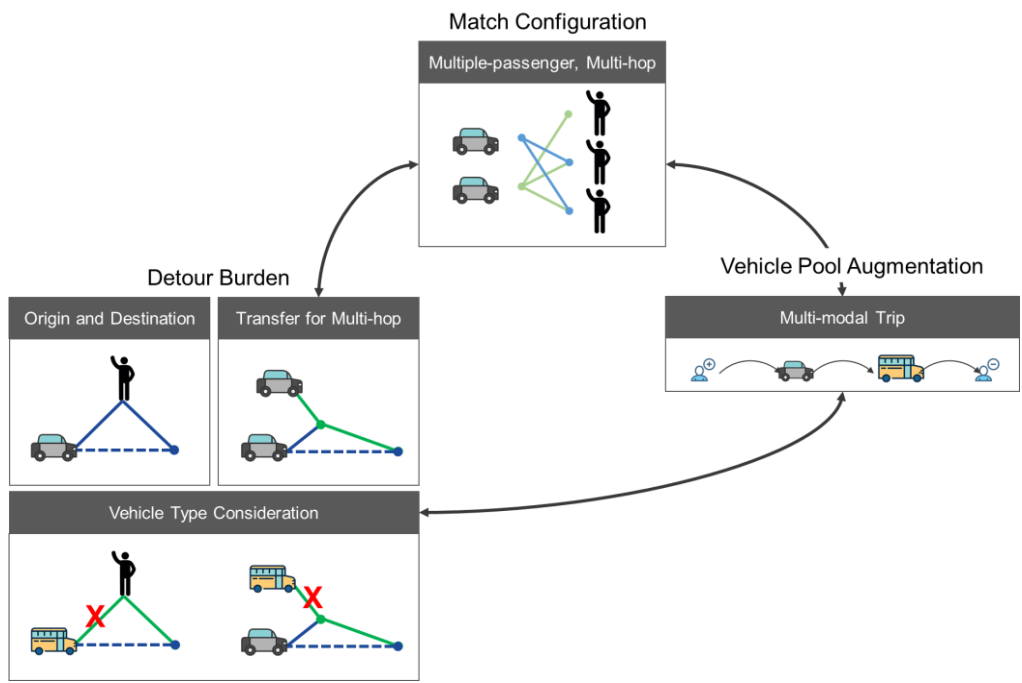

**Figure 4 The Relationship between Three Relaxation Features on Rideshare Matching Algorithm**

Here explains the features for match configuration, detour burden and vehicle pool in detail, regarding the relationship between each other as shown Figure 4. In the match configuration used in this framework, each driver can deliver multiple passengers in a combined travel type as shown in Table 2 and each passenger can take multiple drivers, i.e. multi-hop trips, as shown in Table 3. Maximum transfer for each passenger is one, following a multi-hop ridesharing study that showed a large increase of rideshare efficiency of one transfer, but much smaller marginal gain for more transfers [23]. The vehicle pool is augmented with public transportation available for the rideshare. With multi-hop trips and two modes of transportation, it is possible to permit a multi-modal trip for passengers. Additionally, the detour burden needs to be designed to support the multi-hop and multi-modal trips. Especially with multi-hop, detour burden has to consider the transfer between drivers because their routes may not coincide. Also with multi-modal, detour burden has to consider the vehicle type because public transportation has a fixed route and takes no detour. The different scenarios of detour burden in this framework is organized as Table 4.

**Table 4 The Rideshare Rule for Detour Burden**

| Transportation Mode | For Passenger's Origin | For Passenger's Destination | For Passenger's Transfer |
|---|---|---|---|
| **Private vehicle** | Fully on driver | Fully on driver | - |
| **Public vehicle** | Fully on passenger | Fully on passenger | - |
| **Private to Private** | - | - | Fully on drivers, with negotiation between them |
| **Private to Public** | - | - | Fully on the private vehicle driver |

3.2 Problem Definition

The formal definition of the rideshare matching algorithm is provided in this section. First, the data given for matching is explained. Then, the objective function is shown. The constraints will be explained in more detail in section 3.3.

The information of city geography is given as a graph, $G = \{V, E\}$, where $V$ is the nodes of the city that can be declared as a possible visit, $E$ is the edges between the nodes that represent streets of the city. Calculation of route and travel time is rule-based by the shortest path between two nodes. Congestion is not considered in the travel time.

A request represents entrance of either a participant or a public vehicle to the system. $R_{rider} = \{r_{rider,1}, r_{rider,2}, \dots, r_{rider,i}, \dots, r_{rider,n}\}$ is a set of passenger requests, where $n$ is the number of passenger requests. $r_{rider,i} = \{p_{id,i}, O_{rider,i}, D_{rider,i}, t_{rider,i}^{min,O}, t_{rider,i}^{max,D}, w_i\}$ is a request of passenger $i$, where $p_{id,i}$ is the passenger ID, $O_{rider,i}$ is the origin, $D_{rider,i}$ is the destination, $t_{rider,i}^{min,O}$ is the earliest departure time at origin, $t_{rider,i}^{max,D}$ is the latest arrival time at destination, and $w_i$ is the limit of walking range, respectively of the passenger $i$.

$R_{private} = \{r_{private,1}, r_{private,2}, \dots, r_{private,j}, \dots, r_{private,m}\}$ is a set of private driver requests, where $m$ is the number of private driver requests. $r_{private,j} = \{v_{id,j}, O_{private,j}, D_{private,j}, t_{private,j}^{min,O}, t_{private,j}^{max,D}, c_j\}$ is a request of private driver j, where $v_{id,j}$ is the driver ID, $O_{private,j}$ is the origin, $D_{private,j}$ is the destination, $t_{private,j}^{min,O}$ is the earliest departure time at origin, $t_{private,j}^{max,D}$ is the latest arrival time at destination, and $c_j$ is the seat capacity of the private driver j.

$R_{public} = \{r_{public,1}, r_{public,2}, ..., r_{public,k}, ..., r_{public,b}\}$ is a set of public vehicle requests, where b is the number of public vehicle routes. $r_{public,k} = \{v_{id,k}, S_{public,k}, T_{public,k}, t_{public,k}^{first}, t_{public,k}^{last}\}$ is a request of public vehicle k, where $v_{id,k}$ is the driver ID, $S_{public,k}$ is the visiting station in a fixed order, $T_{public,k}$ is the period of dispatch time, $t_{public,k}^{first}$ is the first dispatch time, $t_{public,k}^{last}$ is the last dispatch time of public vehicle k. The vehicle pool is $D = \{R_{private}, R_{public}\}$ in dimension of (m + b). $v_{id}$ is unique ID given to all driver types including both private and public vehicles. Transfer time is the buffer time to process the transfer from one vehicle to another, given as $t_{transfer}$.

The objective function of this research is to maximize travelers with a given vehicle resources. This projects the assumption of a share system with participants, who are more accepting of the inefficiency of joined trips because they set the travel time window as they can maximally allow for selfish intentions. Therefore, the objective function is formally defined to find a match set M, which maximizes

$$\text{MatchRate}(M) = \frac{Number\ of\ passengers\ in\ match, n^*}{Number\ of\ passenger\ request, n} \times 100\% \qquad (1)$$

3.3 The Rideshare Matching Framework

In this section, a rideshare matching framework is developed to find an efficient set of rideshare matches to solve the rideshare matching problem according to the rideshare rule in section 3.1. After a brief introduction to the rideshare matching framework overall, the following will describe each module in detail.

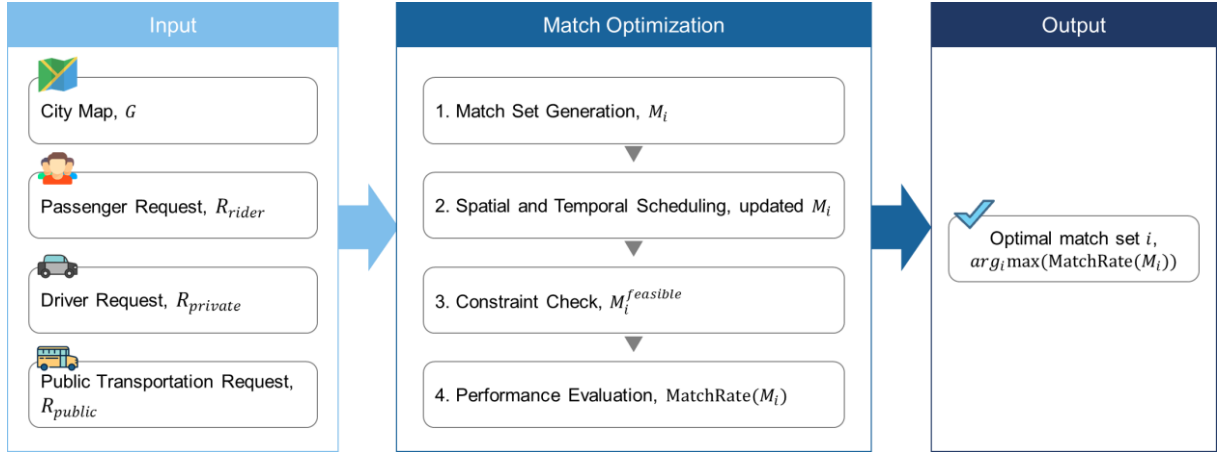

**Figure 5 The Rideshare Matching Framework**

A diagram of the rideshare matching framework is shown in Figure 5. The inputs of the framework includes the city map and rideshare requests as defined in section 3.2. In match optimization, the match set i that maximizes the match rate is found with an optimizing algorithm, via four modules – match set generation, spatial and temporal scheduling, constraint check, and performance evaluation. In match set generation, two drivers are assigned to each and every passenger to produce $M_i$, without time and location of pickup and drop-off yet unknown. In spatial and temporal scheduling, the efficient time and location of pickup and drop-off is calculated for each driver of all passengers in $M_i$ by a rule-based algorithm. In constraint check, the feasible match set $M_i^{feasible}$ is filtered from $M_i$ that meets the constraints of all requests. In performance evaluation, the score of set $M_i$ is calculated. In this paper, the objective function to evaluate is the match rate. The details are explained in the following by modules.

### 3.3.1 Match Set Generation

In this first module, two drivers are assigned for each and every passenger to produce a match set, M. A match set is a set of formed matches between drivers and passengers, given as $M = \{m_1, m_2, …, m_q, …, m_n, m_{n+1}, m_{n+2}, …, m_{2n}\}$ for the multi-hop rideshare with at most one transfer. The proposed rideshare rule is to allow maximum of one transfer, therefore two drivers. $m_q = \{p_{id,q}, order, v_{id,q}, v_{type,q}, x_{pick,q}, x_{drop,q}, t_{pick,q}, t_{drop,q}\}$ is the information of the matched driver for each passenger, where $p_{id}$ is the passenger ID from $R_{rider}$, $order = 1\ for\ q = \{1, …, n\}$ and indicates the first driver of the multi-hop, $order = 2\ for\ q = \{n + 1, …, 2n\}$ and indicates the second driver of the multi-hop, $v_{id}$ is the driver ID from vehicle pool D, $v_{type}$ is the vehicle type of either public or private, $x_{pick}$ is the location of pickup, $x_{drop}$ is the location of drop-off, $t_{pick}$ is the time of pickup, and $t_{drop}$ is the time of drop-off. Naturally for a multi-hop ride, the drop-off of first driver and the pickup of the second driver form the transfer location.

The values of $x_{pick,q}, x_{drop,q}, t_{pick,q}, t_{drop,q}$ for all $m_q$ in $M_i$, q = {1, ... ,2n} are not yet computed in this module. The dimension of the solution space for the optimal match rate is almost $2n^m$, and the computation load exponentially increases with increasing number of driver request, m. Therefore, it may be computationally impractical to find an exact solution for the optimal $M_i$. Rather, meta-heuristic algorithms like Genetic Algorithm may be more practical. The optimization method used in this paper is further discussed in section 3.3.5.

### 3.3.2 Spatial and Temporal Scheduling

In this second module, the rideshare rule in section 3.1 is actively incorporated to calculate the time and location of pickup and drop-off of each driver for all passengers. This module completes and updates $M_i$ with the values of $x_{pick,q}, x_{drop,q}, t_{pick,q}, t_{drop,q}$ for all $m_q$ in $M_i$, q = {1, ... ,2n}. Note that the time and location of transfer is also calculated, as the drop-off of first ride (order=1) and the pickup of second ride (order=2) for $m_q$. The spatial and temporal scheduling follows a rule-based calculation, as following .

3.3.2.1 Spatial Scheduling

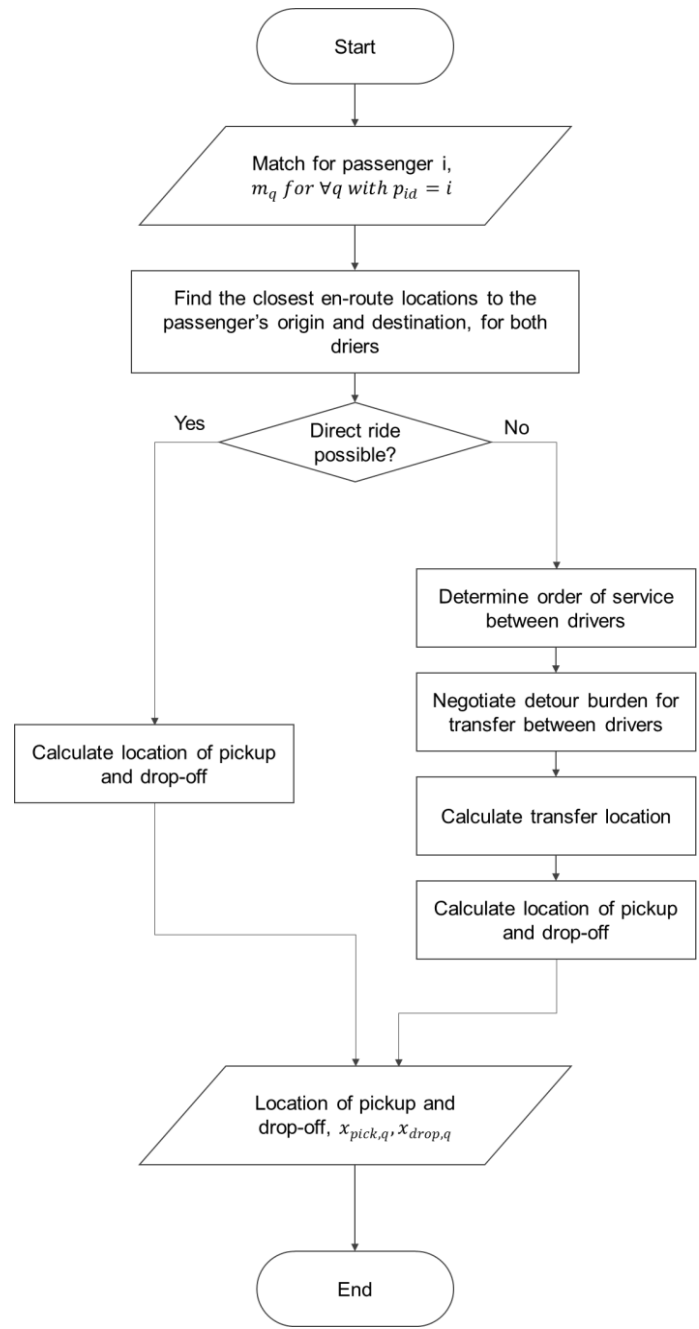

**Figure 6 The Flow Chart of Spatial Scheduling**

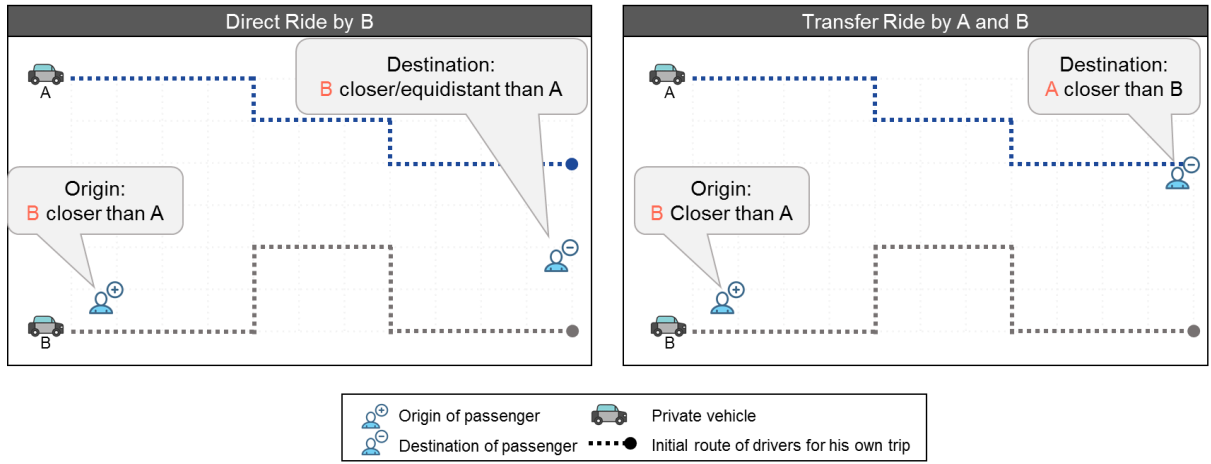

**Figure 7 Determination of Direct Ride**

Spatial scheduling comes before the temporal scheduling. This sub-module calculates the location of pickup and drop-off for each passenger. Its flow chart is shown in Figure 6. Initially, each passenger is assigned two drivers. However, if a direct ride without transfer is possible, only one driver is assigned by dropping the other one from the match. A driver is assigned as a direct ride if he has a closer or equidistant origin to the passenger's origin, as well as a closer or equidistant destination to the passenger's destination. The other driver is no longer necessary and is given null value in the match set. This is exemplified in Figure 7.

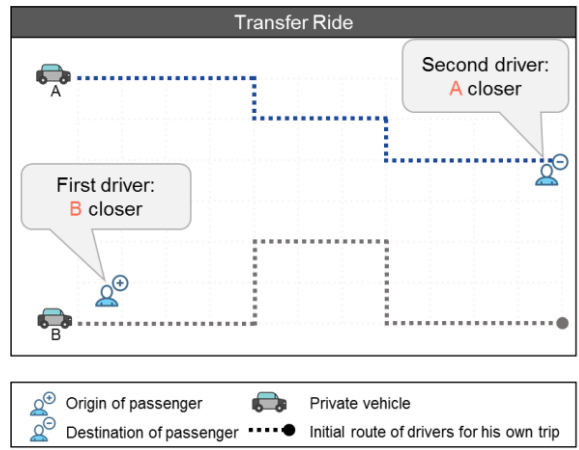

**Figure 8 Order of Service for Transfer Ride**

If a transfer ride is more appropriate, then the order of service between drivers is updated, i.e. who will be the first ride. Initially, the order of service was arbitrarily chosen. However, a driver is updated to be the first ride if he has a closer origin to the passenger's origin. Likewise, the other driver is updated to be the second ride with a closer destination to the passenger's destination. This is exemplified in Figure 8.

After updating the order of service of multi-hop, the detour burden for drivers and passenger is negotiated between drivers for the transfer rides. The detour burden is calculated depending on the vehicle types of the two vehicles and the equation can be written as equations (2) and (3) below.

$$\text{DetourBurden}_{driver}(p_{id}, v_{id}, v_{type}) = 1 - \left(\frac{NumRider(v_{id}, v_{type})}{NumRider(v_{id}, v_{type}) + NumRider(v'_{id}, v'_{type})}\right) \quad (2)$$

Where $\text{DetourBurden}_{driver}(p_{id}, v_{id}, v_{type})$ is the detour burden of driver $v_{id}$ with vehicle type $v_{type}$ for the transfer ride of passenger $p_{id}$ in a ratio, $0 \leq \text{DetourBurden}_{driver} \leq 1$, $NumRider(v_{id}, v_{type})$ is the total number of passengers assigned to driver $v_{id}$ of $v_{type}$, and $v'_{id}$ and $v'_{type}$ are the driver ID and vehicle type of the other driver assigned to $p_{id}$. If $v_{type} = public$, then $NumRider(v_{id}, v_{type}) = \infty$ and Where $\text{DetourBurden}_{driver}(p_{id}, v_{id}, v_{type}) = 0$.

$$\text{DetourBurden}_{rider}(p_{id}) = 1 - \sum_{\forall v_{id} \text{ with } p_{id}} \text{DetourBurden}_{driver}(p_{id}, v_{id}, v_{type}) \quad (3)$$

Where $\text{DetourBurden}_{rider}(p_{id})$ is the detour burden of passenger $p_{id}$, and $\text{DetourBurden}_{rider}(p_{id}) = 1$ if $v_{type} = v'_{type} = public$, in other words two drivers are both public vehicles.

The equation (2) shows the merit-based negotiation of detour burden between two vehicles, where a driver with more passengers to deliver can take less burden in proportion to the passengers that the other driver has to deliver. Also, if the vehicle type is public, the vehicle route is fixed and the detour burden cannot be taken by the vehicle but by the other actors like private driver or passenger. With different set of vehicle types, the scenario of detour burden is varied.

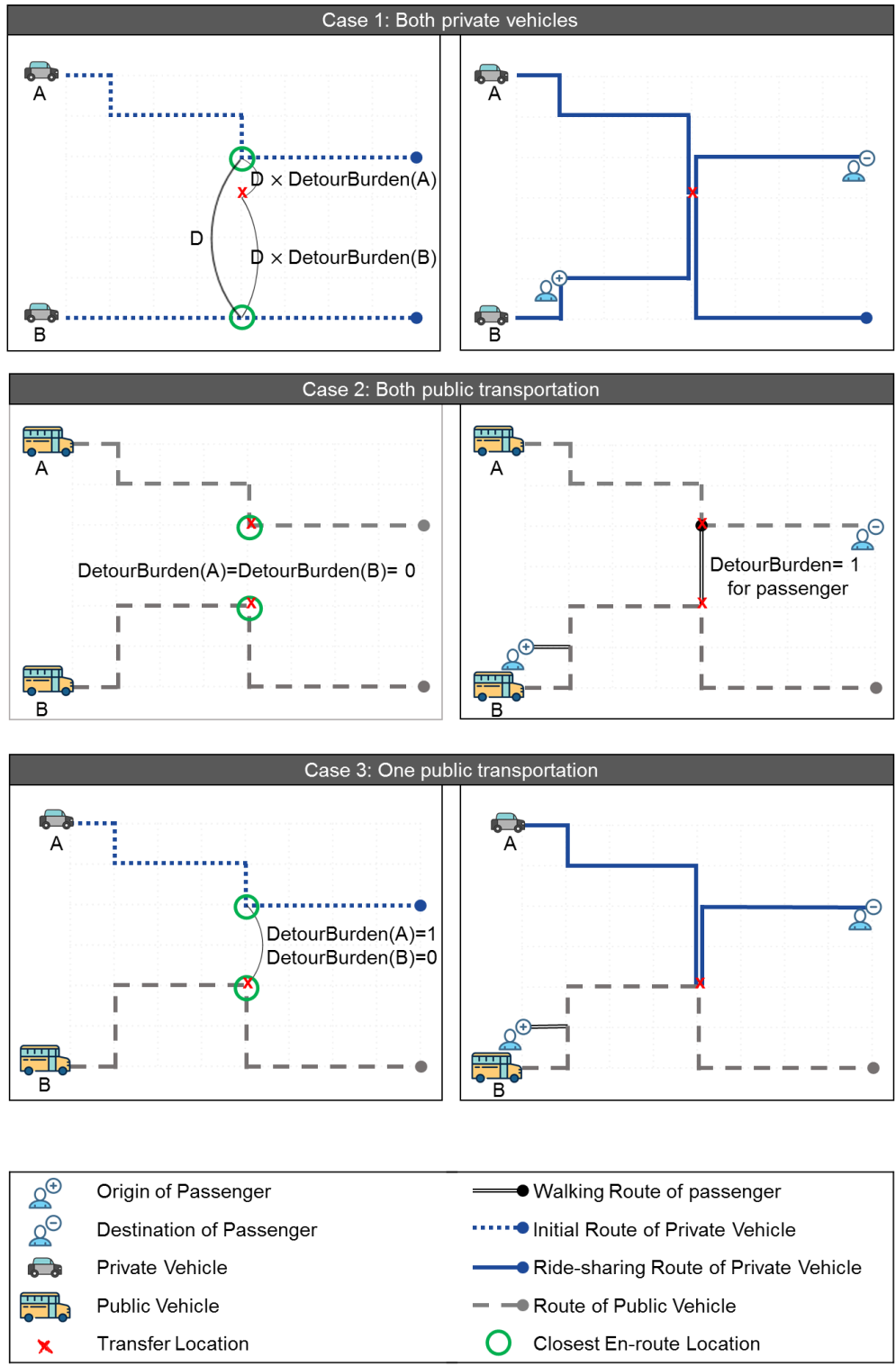

**Figure 9 The Possible Scenarios of Detour Burden for Transfer in Multi-hop**

After the detour negotiation between drivers, the transfer location is calculated based on each driver's detour burden. The routes of the two vehicles for the rideshare are not determined yet. Therefore, the transfer location is calculated between the closest points of the initial routes of the two vehicles for their own journey without ridesharing. The examples of detour burden and transfer location for each set of vehicle types are shown in Figure 9. First, the closest points of the initial routes of the two vehicles are determined as the closest en-route locations. Next, the shortest path between the closest en-route locations is divided in proportion to the detour burden of each vehicle. The division point becomes the transfer location.

Finally for the spatial scheduling, the location of pickup and drop-offs for rest of the rides are calculated, i.e. the pickup and drop-off for the origin and destination of the passengers. Note that the only feasible locations of pickup and drop-off for public vehicles are their fixed list of stations. The location of pickup and drop-off for all types of vehicles for all types of travels are summarized in Table 5.

**Table 5 Location of Pickups and Drop-offs**

| Travel type | Vehicle type | Order of service | Pickup location | Drop-off location |
|---|---|---|---|---|
| **Direct** | Private | - | Passenger's origin | Passenger's destination |
| | Public | - | Closest station to passenger's origin | Closest station to passenger's destination |
| **Transfer** | Private | First ride | Passenger's origin | Transfer location from negotiation |
| | Private | Second ride | Transfer location from negotiation | Passenger's destination |
| | Public | First ride | Closest station to passenger's origin | Closest station to second driver's route |
| | Public | Second ride | Closest station to first driver's route | Closest station to passenger's destination |

### 3.3.2.2 Temporal Scheduling

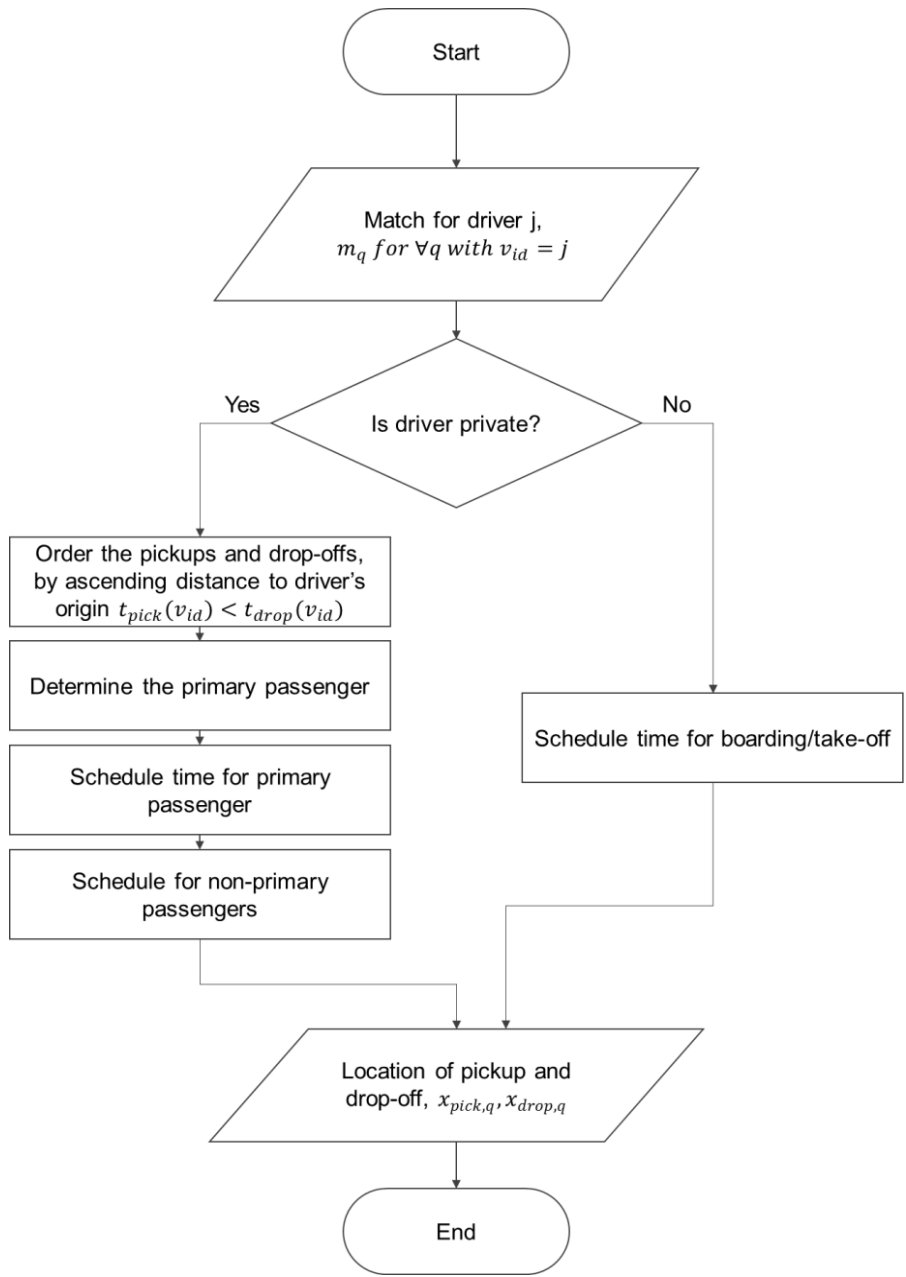

**Figure 10 The Flow Chart of Temporal Scheduling**

With the locations of all visits defined, it is possible to schedule the time of pickups and drop-offs. Similar to the spatial scheduling, the temporal scheduling also depends on the vehicle type. Also to reduce the waiting time of the driver between visits, the time schedule is planned in driver's perspective rather than the passenger's perspective. This means that the time schedule is calculated along the driver's travel route, not passenger's. One absolute rule is that the pickup time is confined to be always earlier than the drop-off time of a driver for any passenger. The flow chart of temporal scheduling is given in Figure 10. Because the time schedule is calculated in driver's perspective, the input for driver $j$ is now $m_q\ for\ \forall q\ with\ v_{id} = j$.

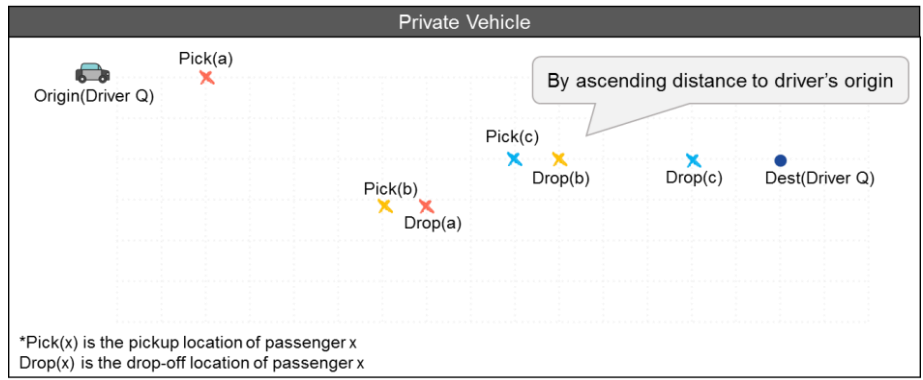

Figure 11 Ordering of Visits for a Private Vehicle

If the driver is a private vehicle, first is to find an efficient order of pickups and drop-offs for the passengers assigned to the driver, using their locations calculated previously. There are some matching algorithms that optimize the rideshare route of vehicles [6], [26]. However, it seems necessary to first test the complex match configuration proposed in this paper with a simple rule of ordering the visits. Shown in Figure 11 is an example of the rule used in this paper, where the locations of pickups and drop-offs are ordered from the closest to the furthest to the driver's origin. The origin and destination of the private driver always come the first and last of the visit order, respectively.

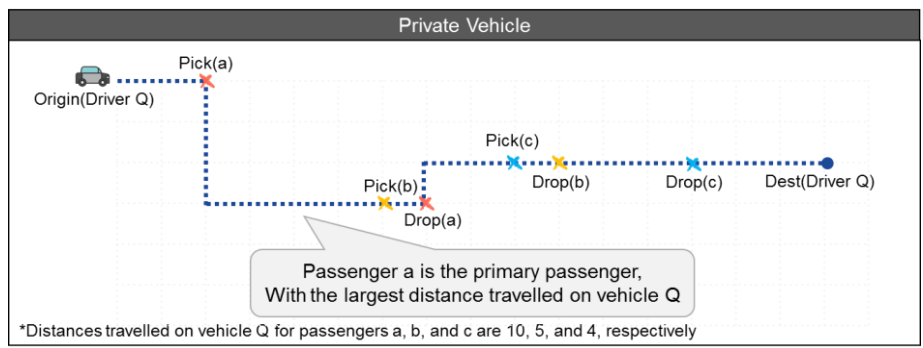

Figure 12 Primary Passenger for a Private Vehicle

Now with the order of visits defined, it is necessary to find a reference point to start the time scheduling. This rule is necessary because time planning for multiple-passenger and multiple-driver configuration is inter-mingled and complicated for instance, a time schedule for pickup of passenger X by driver A affects other drivers the passenger X is assigned to or other passengers the driver A is assigned to. Therefore, a primary passenger can be selected as a reference point to schedule the time of all visits by each driver, i.e. it is the first one among all passengers assigned to one driver to schedule the time for. A primary passenger is chosen as one passenger that travels the longest distance with each driver. The selection of primary passenger for a private driver is exemplified in Figure 12.

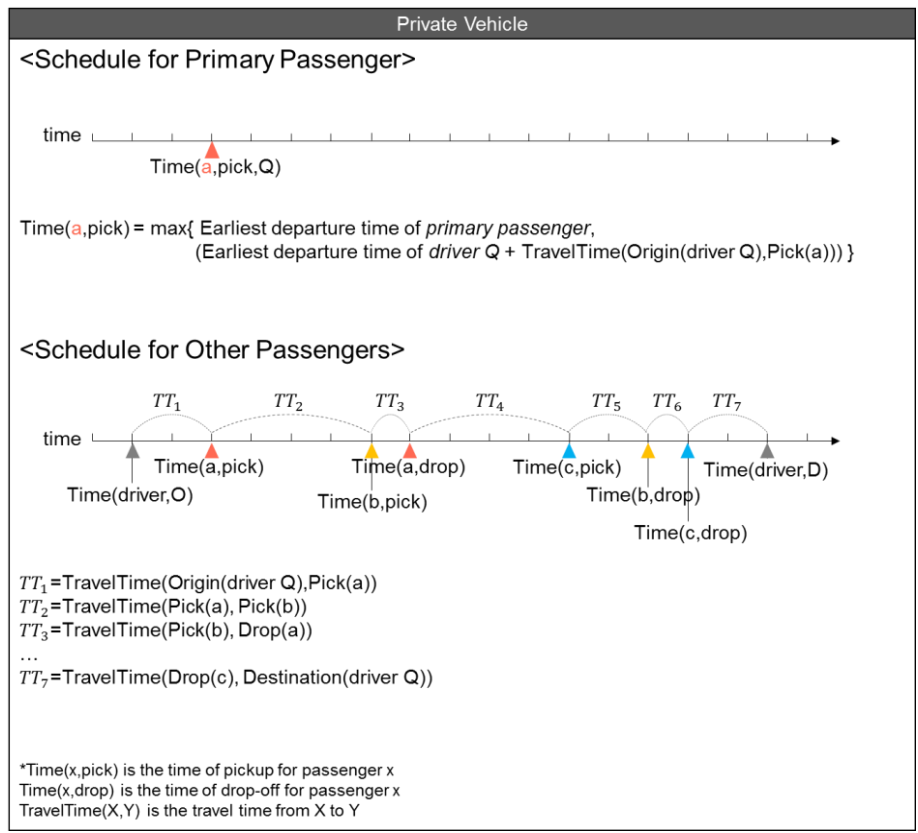

Figure 13 Temporal Scheduling for Primary and Other Passengers

With primary passenger selected, the next step is to schedule time for the primary passenger of the private vehicle as shown in Figure 13. First, only the pickup time for the primary passenger is calculated, with consideration to the earliest departure time of the primary passenger and of the driver. This is the starting point of the temporal schedule for all passengers assigned to the driver. Next, the rest of the pickups and drop-offs is calculated with respect to the order previously defined, with travel time between the ordered visits. Note that the drop-off of the primary passenger may not come immediately after its pickup.

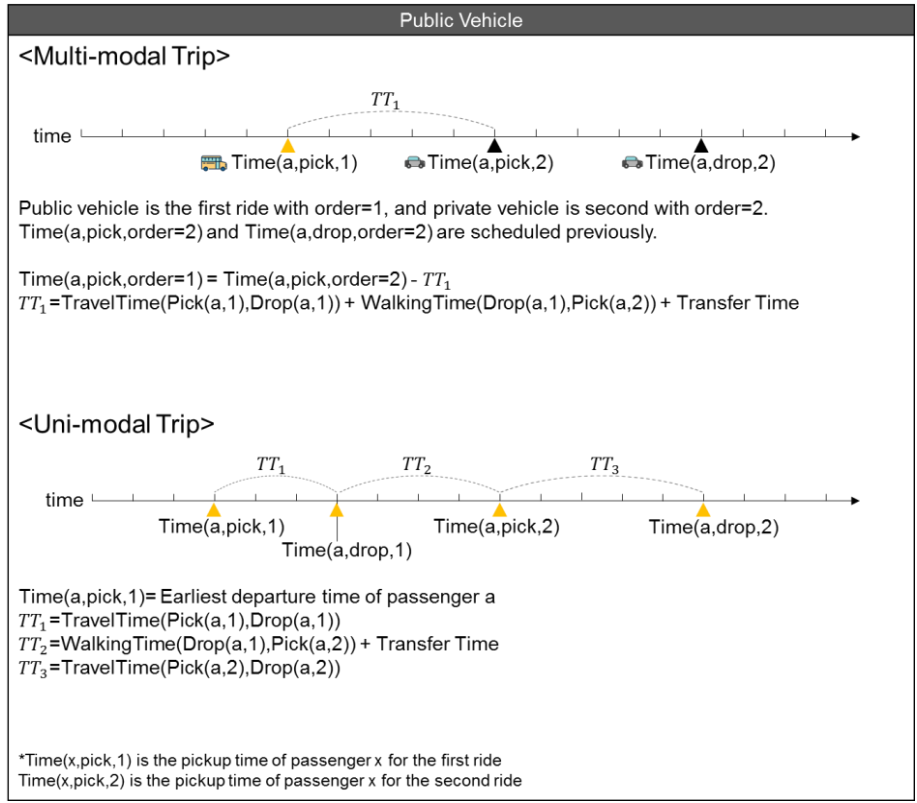

**Figure 14 Temporal Scheduling for Multi-modal and Uni-modal Trips**

If the vehicle is public, temporal scheduling is much simpler because it has a fixed schedule. A periodic dispatch of public vehicle makes the time scheduling only a matter of choosing which schedule is the most efficient for the passenger. The temporal scheduling of passengers using public vehicle is exemplified in Figure 14. For a multi-modal trip, the time of pickup and drop-off of the other private vehicle is the anchor point to schedule the boarding and take-off. For a uni-modal trip, the time of boarding and take-off only depends on the passenger's time window for travel.

### 3.3.3 Constraint Check

In this third module, the constraint of all requests are evaluated with the location and time of all pickup and drop-offs, using a feasible match set. A feasible match set is the subset of a match set that meets the request constraints. Its components are the non-null values of a match set after the constraint check of each $m_q$. It is given as, $M^{feasible} = \{\text{set of } m_q\}$ where $m_q \neq NULL$. Making $m_q$ null is equal to dropping from the match. In the constraint check for all $m_q$ of M $\forall q = \{1, \dots, 2n\}$, $m_q$ is given a null value if,

- $t_{pick,q} < t_{rider,p_{id,q}}^{min,O}$, for the earliest departure time of passenger $p_{id,q}$

- $t_{drop,q} > t_{rider,p_{id,q}}^{max,D}$, for the latest arrival time of passenger $p_{id,q}$

- $t_{origin} < t_{private,v_{id,q}}^{min,O}$, for the earliest departure time of driver $v_{id,q}$ if $v_{type,q} = private$, where $t_{origin}$ is the time to leave the driver's origin to complete the joined trip

- $t_{destination} > t_{private,v_{id,q}}^{max,D}$, for the latest arrival time of driver $v_{id,q}$ if $v_{type,q} = private$, where $t_{destination}$ is the time to arrive at the driver's destination to complete the joined trip

- $t_{pick,q} < t_{public,v_{id,q}}^{first}$, for the earliest departure time of driver $v_{id,q}$ if $v_{type,q} = public$

- $t_{drop,q} > t_{public,v_{id,q}}^{last}$, for the latest arrival time of driver $v_{id,q}$ if $v_{type,q} = public$

Also, $m_q = NULL$ for all with $v_{id} = v_{id,q}$ if $v_{type,q} = private$ and if

- $c_{v_{id,q}} <$ (Number of match $m_p$'s that shares the same $v_{id,q}$), for the seat capacity of vehicles.

Similarly, $m_q = NULL$ for all with $p_{id} = p_{id,q}$ if,

- $w_{p_{id,q}} <$ (Distance from $x_{drop}$ of order=1, to $x_{pick}$ of order=2), for limit of walking range of riders.

- ($t_{pick}$ of order=2) < ($t_{drop}$ of order=1) $+ t_{transfer}$, for the time for the next pickup in the multi-hop

### 3.3.4 Performance Evaluation

In this forth module, the performance of match $M_i$ is measured. The match rate is used as a measure, shown with equation (1) in Problem Definition. This value is recorded and saved with $M_i$ to compare with the performance with other $M_i$'s. Note that because the spatial and temporal scheduling is a rule-based algorithm, the $M_i^{feasible}$ and the match rate are always the same for a given match set, $M_i$.

### 3.3.5 Match Optimization

As mentioned in section 3.3.1, the solution space for the match set $M_i$ is too large as $2^{n^m}$, to find an exact solution. Instead a meta-heuristic algorithm can be used to find a match set that maximizes the match rate in a reasonable computing time. Genetic Algorithm (GA) is one of the often used meta-heuristic algorithm for rideshare matching problem with sufficient results [7], [8], [10], [20]. Genetic Algorithm mimics the process of natural selection, where a population of chromosomes is iteratively produced by modifying the previous population's chromosomes with high fitness. Though it may not be the optimal, it allows fast computation for the problem with exponentially increasing computation load and finds a solution of which the performance converges at a level. It is possible to increase the converging fitness by adjusting the parameters and detailed algorithms, such as generation of initial population.

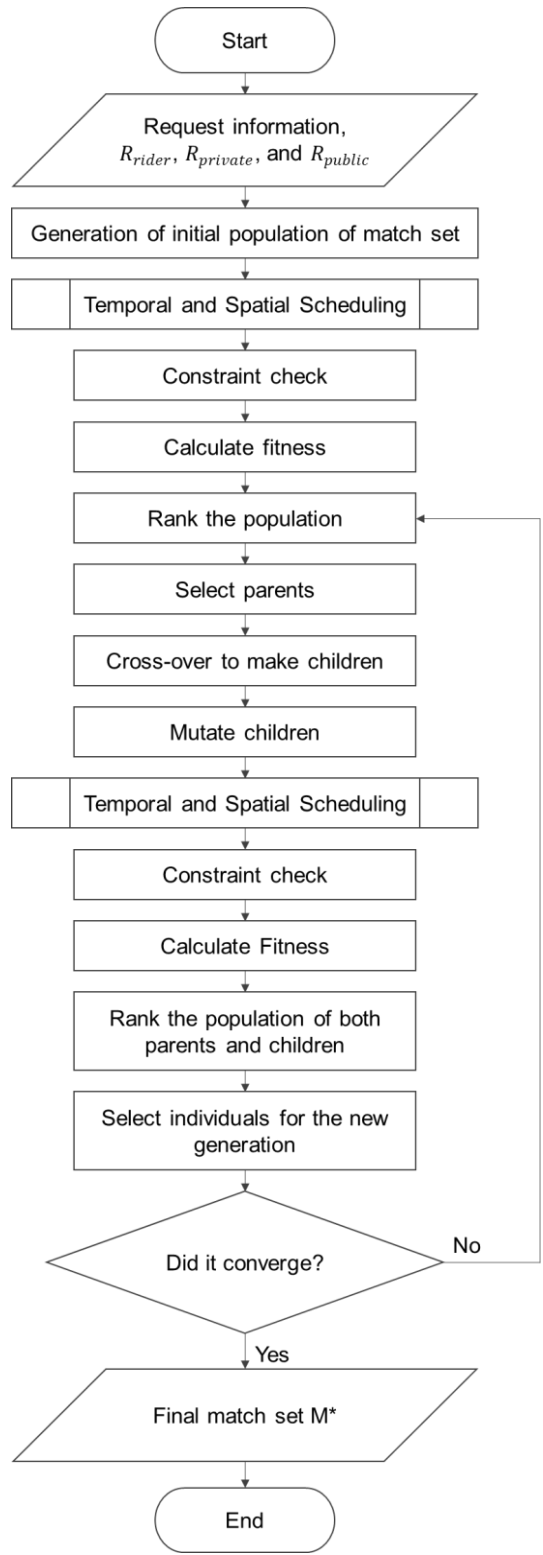

**Figure 15 Flow Chart of the Genetic Algorithm for the Rideshare Matching Framework**

To use Genetic Algorithm for the rideshare matching problem, the four modules previous discussed in section 3.3.1 to 3.3.4 are fitted to the GA as shown in Figure 15. The first module of match generation in section 3.3.1 is executed in initial generation of match set of GA. Then, the second module of spatial and temporal scheduling in section 3.3.2 is executed as a pre-process step before the constraint check. Following is the third module of constraint check in section 3.3.3, which is also the constraint check of GA. Fitness calculation in the GA is the forth module of performance evaluation as discussed in section 3.3.4. The final output of the GA is the match set with the best fitness under the GA setting.

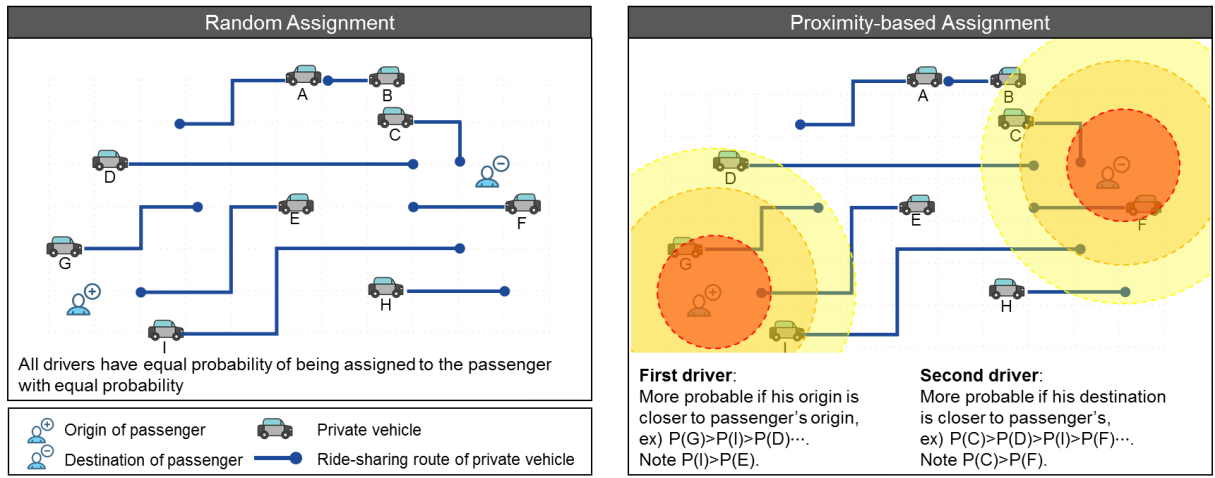

Figure 16 Modification of Initial Population Generation

Additionally, the performance of GA can be enhanced by modifying the initial generation of population and increasing the initial fitness value. Instead of random selection of drivers to passengers, a proximity-based assignment can be used shown in Figure 16. This assignment method gives more probability for participants to be matched in initial population if they are close to each other. The probability of a driver being assigned to a passenger is be increased if his origin is closer to the passenger's origin or if his destination is closer to the passenger's destination. It is possible to give merits to the drivers with closer origin and closer destination separately, since at most two drivers for each passenger need to complete the passenger's journey by providing each leg to the origin and destination.

Finally, the pseudo-code for the rideshare matching framework using Genetic Algorithm is shown below in Table 6.

**Table 6 The Pseudo-code for the Rideshare Matching Framework**

| **Input** |
|---|
| *Rideshare Request Input* |

- Number of passenger requests, n
- Number of private driver requests, m
- Number of public driver requests, b
- Passenger request set, $R_{rider}$
- Private driver request set, $R_{private}$
- Public driver request set, $R_{public}$
- Vehicle pool, D
- City network, G

*Ride-share Parameters*
- Transfer Time, $t_{transfer}$

*Genetic Algorithm Parameters*
- Size of population, Q
- Mutation Rate, $r_{mutate}$
- Cross-over rate, $r_{cross-over}$
- Convergence Threshold, $C_{lim}$

**Output**
Maximum fitness of the last population, $f'_{\max(IterNum)}$
Match set with the maximum fitness, $M_{\max(IterNum),l}$ $where$ $f_{\max(IterNum),l} = f'_{\max(IterNum)}$

**Code**
  **#Module 1: Initial Generation**
  1. For each individual $M_{0,l}$ in initial population $P_0$, *l=1…Q* {
      1. For each passenger request i in $R_{rider}$, $i = 1, …, n$ {
          1. Proximity-based assignment of two drivers from vehicle pool, D.}}
  2. *IterNum*=0.
  3. Save the initial population of match set, $P_0$.

  4. For each individual $M_{0,l}$ in initial population $P_0$, *l=1…Q* {
      **#Module 2: Spatial and Temporal Scheduling of Rides**
      #Module 2.1: Spatial Scheduling with Burden Assignment
      2. For each passenger ID i in $M_{0,l}$, $i = 1, …, n$ {
          2. Calculate closest en-route locations to the passenger's origin and destination for the two assigned drivers
          3. If direct ride is possible {
              2. Drop the other ride from passenger i.}
          4. Else if transfer ride {
              3. Determine order of service between the two rides for passenger $i$
              4. Negotiate the detour burden between two rides
              5. Calculate the connecting location}
          5. Update pickup and drop-off locations for all drivers.}

      #Module 2.2: Temporal Scheduling for Private Vehicles
      3. For each private driver ID j in $M_{0,l}$, $j = 1, …, m$ {
          6. Order the pickups and drop-offs. The constraint is that each passenger's pickup comes before drop-off.
          7. Calculate the time schedule of pickups/drop-offs for primary passenger of driver j
          8. Calculate the time schedule of pickups/drop-offs for all other passengers of driver j }

      #Module 2.3: Temporal Scheduling with Public Vehicles
      4. For each passenger ID i in $M_{0,l}$, $i = 1, …, n$ {
          9. If passenger i is assigned bus, calculate the time schedule of boarding and leaving of the bus.}

      **#Module 3: Constraint Check**
      #Module 3.1: Private Vehicles
      5. For each private driver ID j in $M_{0,l}$, $j = 1, …, m$ {

10. If any visit of private driver j is before the earliest departure time or after latest arrival time {
   6. Drop all passengers she's assigned to.}
11. Else if private driver $j$ is assigned more than seat capacity {
   7. Drop all passengers she's assigned to.}

#Module 3.2: Public Vehicles
6. For each passenger ID i in $M_{0,l}$, $i = 1, ..., n$ {
   12. If passenger i is assigned bus {
      8. If the pickup time is earlier than the bus's first service time {
         1. Drop all rides assigned to passenger $i$.}
      9. Else if the drop-off time is later than the bus's last service time {
         2. Drop all rides assigned to passenger $i$.}

#Module 3.3: Passengers
7. For each passenger ID i in $M_{0,l}$, $i = 1, ..., n$ {
   13. If the first pickup time is before her earliest departure time {
      10. Drop all rides assigned to passenger $i$}
   14. Else if the last drop-off time is after her latest arrival time {
      11. Drop all rides assigned to passenger $i$.}
   15. Else if transfer ride and if the second pick-up is earlier than the first drop-off with $t_{transfer}$ and walking time {
      12. Drop all rides assigned to passenger $i$.}

#Module 4: Fitness Calculation
8. Calculate the fitness of each individual $M_{0,l}$, $f_{0,l}$.
}
5. Save the maximum fitness of $P_0$, $f'_0$

6. $C_{converngece}=0$
7. While $C_{converngece} < C_{lim}$ {
   9. $IterNum = IterNum + 1$

#Module 5: Making New Generation
10. For each individual $M^{child}_{IterNum,l}$ in child population $P^{child}_{IterNum}$, $l=1...Q$ {
    16. Select two parents $M_{mother}$ and $M_{father}$ of previous population, $P_{IterNum-1}$ by Roulette Wheel.
    17. Cross-over the parents with $r_{cross-over}$ to make a child, $M_{child}$.
    18. Mutate each string of $M_{child}$ with $r_{mutate}$.
    19. $M^{child}_{IterNum,l} = M_{child}$.
    20. Execute Module 2 on $M^{child}_{IterNum,l}$
    21. Execute Module 3 on $M^{child}_{IterNum,l}$
    22. Execute Module 4 on $M^{child}_{IterNum,l}$ to produce the fitness $M^{child}_{IterNum,l}$, $f^{child}_{IterNum,l}$}
11. Save the child population, $P^{child}_{IterNum}$

12. For each individual $M_{IterNum,l}$ in adult population $P_{IterNum}$, $l=1...Q$ {
    23. If $f^{child}_{IterNum,l} < f_{IterNum-1,l}$ {
       13. $M_{IterNum,l}=M_{IterNum-1,l}$}
    24. Else {$M_{IterNum,l}=M^{child}_{IterNum,l}$}
}
13. Save the new population, $P_{IterNum}$.
14. Save the maximum fitness of $P_{IterNum}$, $f'_{IterNum}$

#Module 6: Convergence Check

> 15. $f'_{IterNum} - f'_{IterNum-1} 0.00001\{$
>      24. $C_{converngece}=C_{converngece}+1$. If not, $C_{converngece}=0\}$
> }
>
> 8. Return the last population, $P_{max(IterNum)}$
> 9. Return the maximum fitness of the last population, $f'_{max(IterNum)}$
> 10. Return the match set with the maximum fitness, $M_{max(IterNum),l}$ where $f_{max(IterNum),l} = f'_{max(IterNum)}$

### 3.4 Verification

#### 3.4.1 Methodology

There are two verifications needed for the performance of the proposed framework. First is to verify that each rideshare match completes the trip of participants, using the route examples of matches found. Second is to verify that the Genetic Algorithm performs appropriately. A sample of converged fitness is given together with computation time. Both verification methods take a simplified network and data generated with its simplified origin-destination trip data.

#### 3.4.2 Data

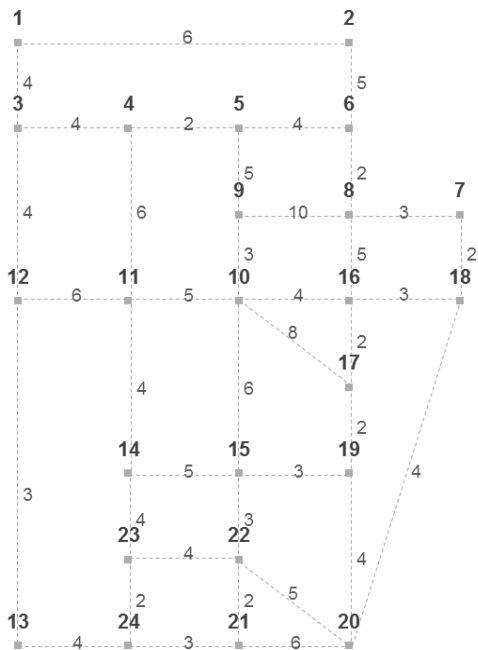

**Figure 17 The Simplified Network of Sioux Falls**

The data used in the verification of the proposed framework is a simplified network of Sioux Falls, which has 24 nodes and 76 links. The links represent the streets, which are assumed all bi-directional with the same free flow speed. The map of the network is given Figure 17, with bold numbers indicating the node name. The labels on the links indicate the free flow travel time of the link. The rideshare matching parameters are 5 minutes for the transfer time and 10 minutes for the limit of walking. The earliest departure time is randomly generated from 9AM to 10AM for all participants. The temporal flexibility is 40 minutes for the route examples, 20 minutes for fitness convergence of GA, and 30 minutes for computation time of GA. The ratio of vehicle supply to demand is 200% for route examples and fitness convergence and 250% for computation time of GA, with no public transportation.

For the Genetic Algorithm, the stopping criteria is the maximum number of iteration with unchanging fitness value, i.e. convergence threshold. The convergence threshold is 30 iterations for the first verification method, the route examples. The convergence threshold is 500 iterations for the second verification method, the fitness convergence of GA. Other parameters of GA are population size of 20, mutation rate of 0.01, and cross-over rate 1 with single-point cross-over at random sampled location from uniform distribution along the chromosome span, and fitness variable as the match rate. The objective function is to maximize the match rate. The framework is computed using a software R, with Intel Core i5-4670 CPU @ 3.4GHz with RAM 16.0GB.

### 3.4.3 Results

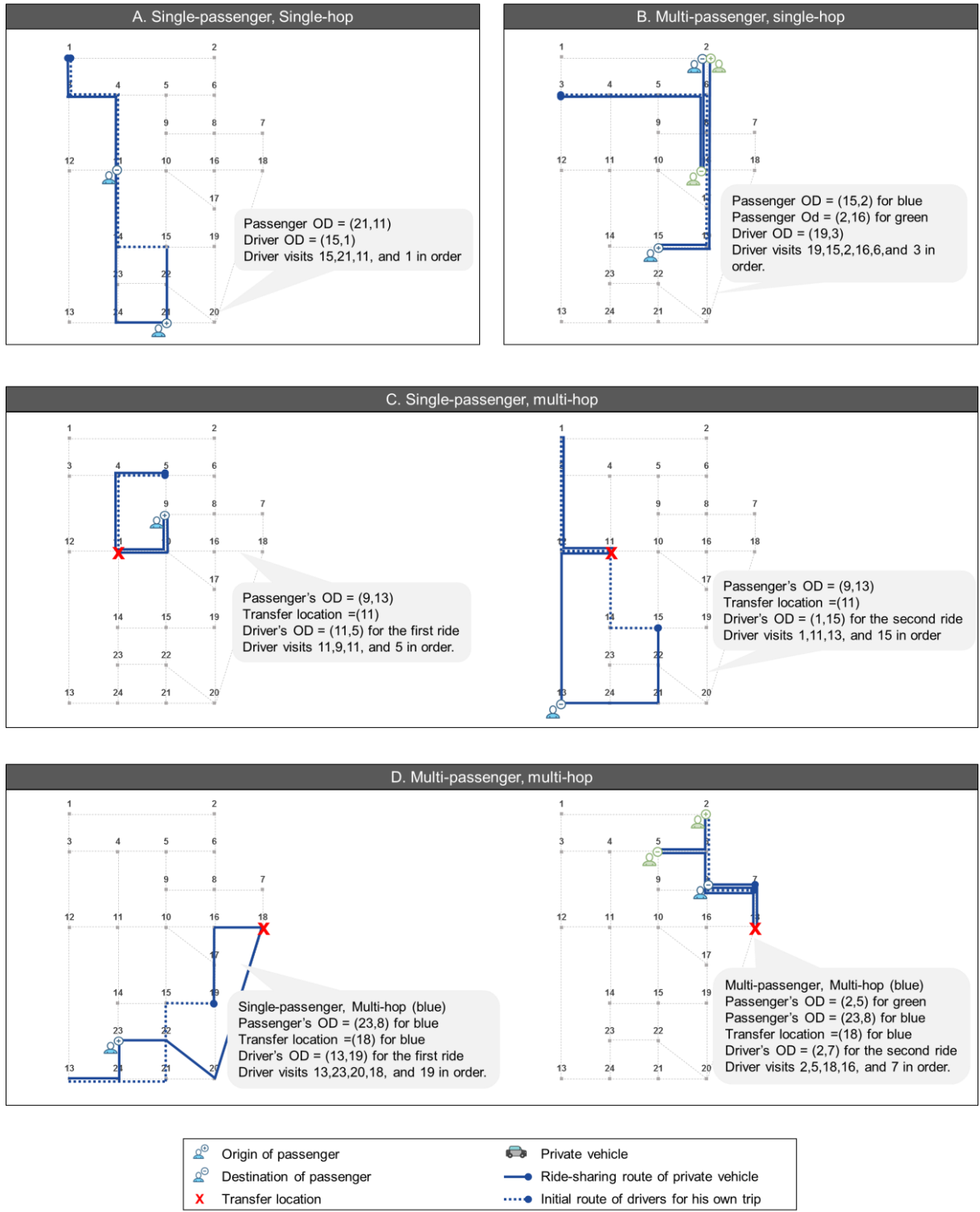

**Figure 18 Examples of Routes for Matched Rideshare**

First to verify the performance of the proposed framework, the route examples are shown where the formed matches complete the trip of participants in Figure 18. The participants' origin and destination (OD) is indicated in the figure, as well as the transfer locations. In Figure 18-A, an example of a rideshare match between a single passenger and single driver is shown to complete both of their trips with some detour. In Figure 18-B, an example is shown, where the driver detours to finish trips of two passengers, blue and green, and his own. In Figure 18-C, an example of multi-hop is shown of a passenger, whose drivers both take only one passenger. Finally, in Figure 18-D, an example of multi-hop is shown of a passenger, of which one of the drivers takes multiple passengers. The route of each driver is calculated based on rules where the shortest paths between their pre-ordered visits are taken to complete the travel. Therefore, the route may not be optimal for the drivers.

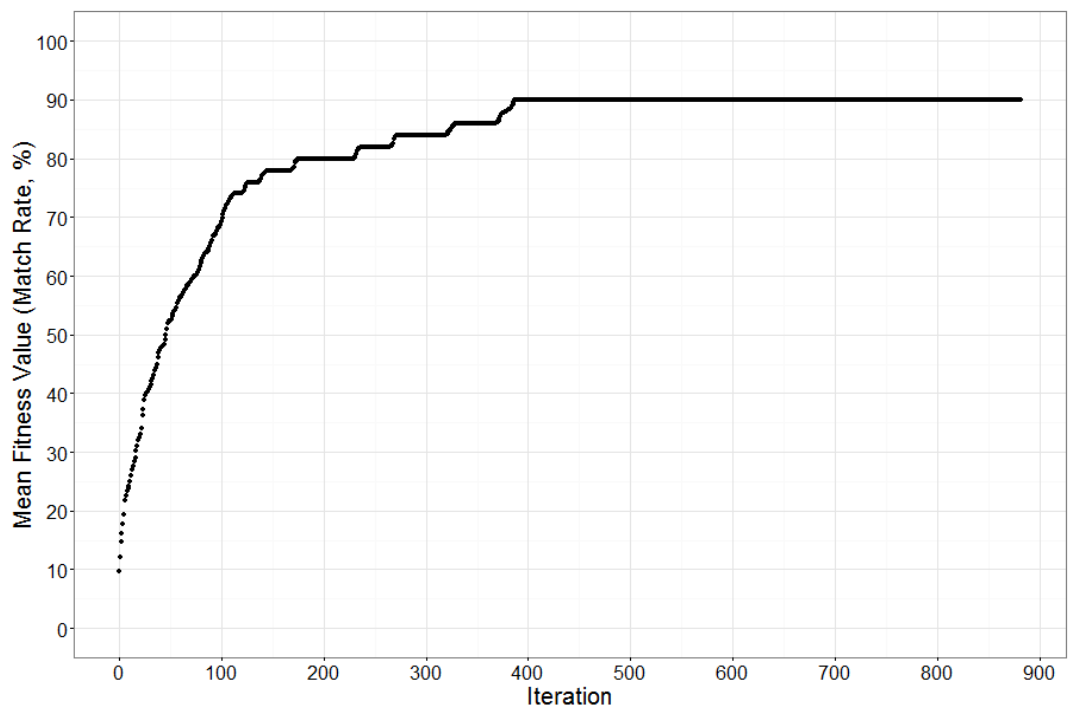

**Figure 19 Fitness Convergence of the Proposed Framework**

Second to verify the performance of the framework with the Genetic Algorithm, Figure 19 is shown. Convergence of the GA fitness is shown with a steady match rate value that has reached about 90%. The convergence threshold, or the 'run' of the GA, is used as the stopping criteria. With smaller convergence threshold, it gives the algorithm a less chance to escape local optima to find a global optimal, thereby reduce the maximum fitness. However, it may also largely reduce the computation time by finishing the search with less iterations. Therefore, the convergence threshold is used much less with 30 iterations in the later chapters, to reduce the computation time.

Additionally, to verify a practical computation time of the framework, 30 samples are reproduced for statistical results. With passenger request of 50 and driver request of 125 with no public transportation, the computation mean is 14.2 minutes with standard deviation of 2.9 minutes. The mean match rate is 80.8% with standard deviation of 5.3%.

3.5  Conclusion

In this chapter, a rideshare matching framework is developed that allows the most advanced form of sharing mobility with multi-modal trips, in order to maximize the matching efficiency. In section 3.1, the rideshare rule is carefully designed with consideration to the relationship between match configuration, detour burden, and vehicle pool augmentation, whose potential is discussed in the literature review of Chapter 2. Section 3.2 formally defines the rideshare matching problem and section 3.3 gives the proposed framework for the rideshare matching using a meta-heuristic method, Genetic Algorithm. The proposed framework is verified in section 3.4

The uniqueness of the developed rideshare matching framework is that it allows multi-modal trips in the most flexible form of rideshare matching, i.e. multi-hop and multi-passenger form, with public vehicles. It may help increase the probability of successful matching with unequal size of driver and rider requests, therefore inviting more participants to achieve a critical mass of stable match rate. For this advancement, three relaxation features are considered in building the rideshare rule and matching algorithm. For match configuration, multiple-passenger and multi-hop are allowed. For vehicle pool augmentation, public transportation is integrated in the matching algorithm as an additional vehicle option. For detour burden design, more realistic burden negotiation rule between the participants is developed to incorporate multi-hop and additional vehicle pool in a match.

A meta-heuristic method is used for maximization of match rate in the framework, since an exact solution with an exponentially increasing computation burden is impractical. Despite the complex rideshare rules, it is very simple to integrate the rideshare rules and constraints to the flow of GA as modules. The verification results show examples of successful completion of a joined trip for various match types, as well as successful performance of GA with reasonable computation time. In the next two chapters, the developed framework will be used to evaluate the effects of additional vehicle pool and the effects of schedule flexibility of participants on the rideshare matching efficiency

# Chapter 4. Public Transportation in Rideshare

In this chapter, the effect of public transportation is evaluated with the performance of the multi-modal rideshare matching framework. As one of the relaxation features of the rideshare matching, public transportation is added to the vehicle pool to increase the matching efficiency even at an unequal size of drivers and passengers. To provide useful insight for practical application, it is necessary to evaluate the improvement of matching efficiency with this feature at different levels of rideshare vehicle supply. The following sections describe the methodology, data, and results of the analysis. This chapter is finished with a conclusion.

## 4.1 Methodology

In order to test a positive influence of additional vehicle pool on matching efficiency, it needs to be shown that more public vehicles increase the matching efficiency. Therefore, the period of dispatching schedule of public vehicles is used as a measure to represented the vehicle flow in a city at a given time. The larger the period of public vehicle dispatch, the smaller the flow of public vehicles in the system. The match rate of a given set of requests will be evaluated for different dispatch period of public vehicles, for different levels of private vehicle supply.

The terms used in this paper for evaluation are defined here. Ratio of supply to demand (RSD) describes the level of supply for a given demand, as equation (4),

$$Ratio\ of\ Supply\ to\ Demand\ (RSD) = \frac{Number\ of\ driver\ request, m}{Number\ of\ passenger\ request, n} \times 100\% \qquad (4)$$

Note that RSD counts only the private vehicles but does not include public vehicle as it is a given parameter to a matching environment. Dispatch period of bus is given in minutes and each bus route represents a bi-directional route of a fixed dispatch period.

## 4.2 Data

The network used in this chapter is same to Figure 17 in Chapter 3, the simplified network of Sioux Falls. The rideshare matching parameters are 5 minutes for the transfer time, 20 minutes for the limit of walking, number of passenger request of 100, and RSD of 10,50, and 100%. The earliest departure time is randomly generated from 9AM to 10AM for all participants. The temporal flexibility for participants is randomly generated with uniform distribution of 0 to 60 minutes. Sample size is described for each figure of results. The seat capacity of each driver is 3 passengers and each passenger can take only one transfer, making two hops. For the Genetic Algorithm, the convergence threshold is 30 iterations, population size of 20, mutation rate of 0.01, and cross-over rate of 1 with single-point cross-over at random sampled location from uniform distribution along the chromosome span, and fitness variable as the match rate with objective function to maximize it.

The bus data used in this paper are generated to cover the streets of the entire network. The real bus routes of Sioux Falls are first simplified to fit the network and are augmented both in number of routes and lengths to touch most of the streets of the simplified network. The total bus routes are shown on the map of the network as well as in table in Figure 20 and Table 7. The period of bus dispatch used is 1, 5, and 30 minutes, which represents 713, 143, and 24 public vehicles per minute in the city. Route ID of the same number is counted as one route in the table, summing up to total of 9 routes of public transportation generated.

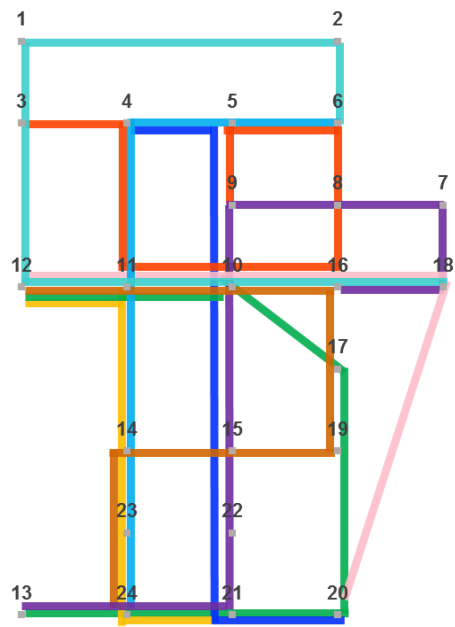

**Figure 20 Bus Routes for the Simplified Network**

(* The ID and route information of the bus routes are given in Table 7)

**Table 7 Route Information of the Bus**

| Route ID | Route color | Origin | Destination | Visiting stations (with direction) |
|---|---|---|---|---|
| 1A | Yellow | 12 | 21 | 12,11,14,23,24,21 |
| 2A | Red | 3 | 9 | 3,4,11,10,16,8,6,5,9 |
| 3A | Blue | 4 | 20 | 4,5,9,10,15,22,21,20 |
| 4A | Purple | 13 | 16 | 13,24,21,22,15,10,9,8,7,18,16 |
| 5A | Green | 12 | 13 | 12,11,10,17,19,20,21,24,13 |
| 6A | Cyan | 6 | 18 | 6,2,1,3,12,11,10,16,18 |
| 7A | Light Blue | 6 | 24 | 6,5,4,11,14,23,24 |
| 8A | Pink | 12 | 20 | 12,11,10,16,18,20 |
| 9A | Brown | 12 | 24 | 12,11,10,16,17,19,15,14,23,24 |
| 1B | Yellow | 21 | 12 | 21,24,23,14,11,12 |
| 2B | Red | 9 | 3 | 9,5,6,8,16,10,11,4,3 |
| 3B | Blue | 20 | 4 | 20,21,22,15,10,9,5,4 |
| 4B | Purple | 16 | 13 | 16,18,7,8,9,10,15,22,21,24,13 |
| 5B | Green | 13 | 12 | 13,24,21,20,19,17,10,11,12 |
| 6B | Cyan | 18 | 6 | 18,16,10,11,12,3,1,2,6 |
| 7B | Light Blue | 24 | 6 | 24,23,14,11,4,5,6 |
| 8B | Pink | 20 | 12 | 20,18,16,10,11,12 |
| 9B | Brown | 24 | 12 | 24,23,14,15,19,17,16,10,11,12 |

4.3 Results

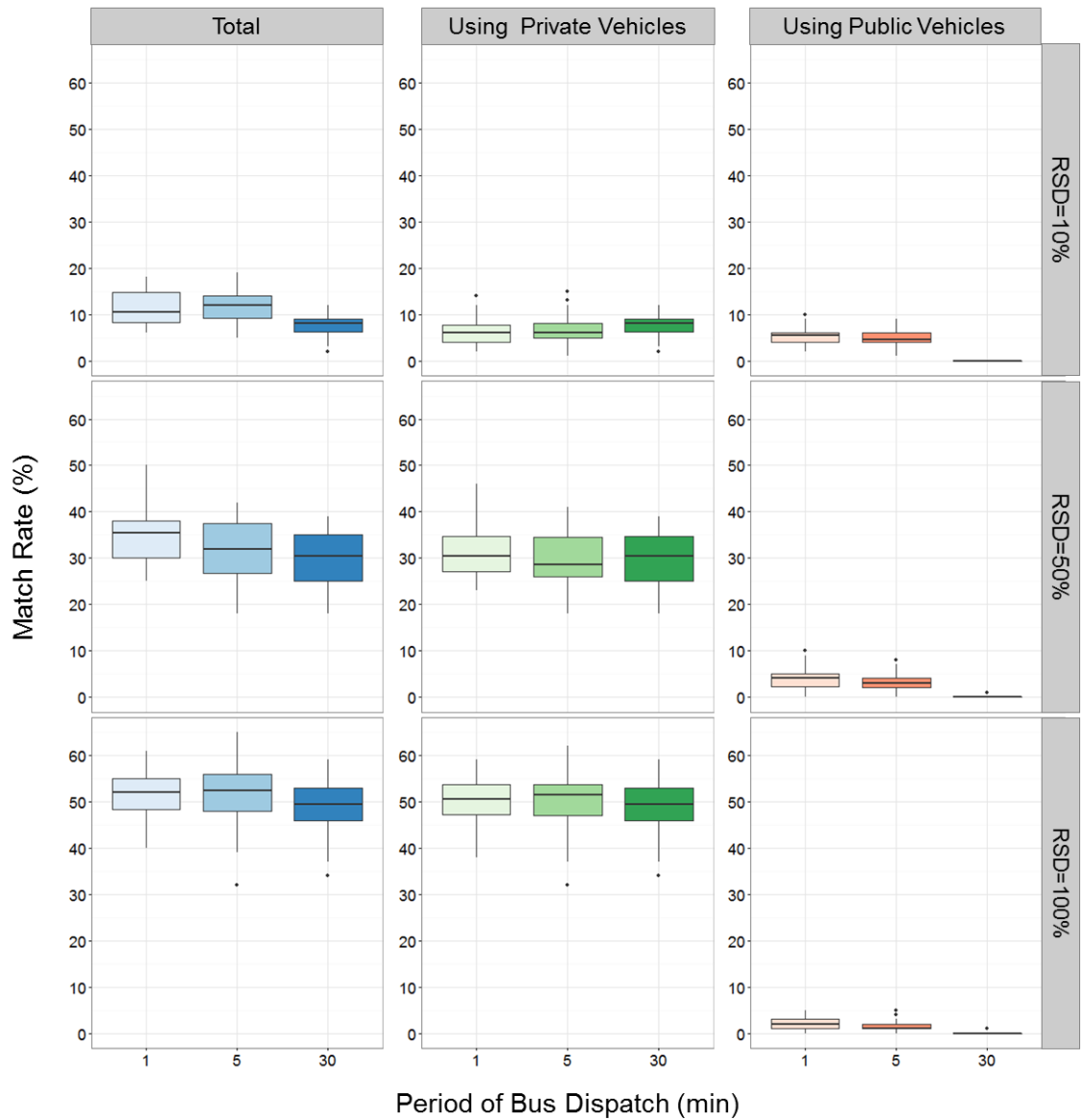

**Figure 21 Match Rate of Increasing Period of Bus Dispatch for Different Ratios of Supply to Demand**

The result of this chapter is shown in Figure 21. There are nine graphs, all with x-axis as increasing period of bus dispatch, i.e. decreasing flow of public vehicle at a time, and y-axis of maximum match rate reached in each rideshare matching sample of GA. Note x-axis is not to scale. The graph columns indicate the match rate as equation (1), the match rate for private vehicles, and the match rate for public vehicles, from the left to right respectively. The match rate for private vehicles is the number of passengers matched with private vehicles for a rideshare divided by the total number of requests, n. The match rate for public vehicles is the number of passengers matched with public vehicles for a rideshare divided by the total number of requests. This is regardless of whether the passenger takes a multi-hop trip or not. If a passenger travels in both private and public vehicles, the match is counted both in the match rate for private and public vehicle. Therefore, the match rates of second and third columns sum equal or larger than the match rates of first column. The graph rows have different Ratio of Demand to Supply values of 10, 50 and 100% from top to bottom, respectively. The sample size for each dispatch period and each RSD value is 30, making up total samples of Sample size: 30 samples×3 dispatch periods ×3 RSD values=270 samples.

First from the figure, it is evident that as Ratio of Demand to Supply increases, the match rate generally increases regardless of additional supply of public transportation. For instance, the match rate for RSD of 10% is around 10%, whereas the match rate for RSD of 50% is around 30 to 35%. More private vehicles satisfy more demand, but not all vehicles may be used for the rideshare. For example, even when the supply size is the same as the demand size at RSD of 100%, the match rate is less than 100%.

Second, the impact of public transportation on the match rate is analyzed. With very few private drivers at RSD of 10%, larger dispatch period of public vehicles slightly influences negatively to the total match rate, i.e. larger supply of public vehicles slightly influences positively to the total match rate. This is shown by the increasing trend of match rate in the total match rate on the left. The total match rate decreases because the match rate for public vehicles decreases. The discrepancy between the match rates from private vehicles and public vehicles is not too large at this stage.

As the supply of private driver increases with higher RSD, the effect of more public vehicle on match rate subsides even more. For instance, it is difficult to observe for RSD of 50% the decreasing trend of match rate with larger dispatch period and smaller public vehicle flow. The reason is the match rate from private vehicle largely increased with RSD so that the match rate from public vehicles is relatively much smaller to show its effect. For instance, at RSD of 100%, the match rate ranges from 0 to 5% for public vehicle and around 50% for private vehicle. When the supply of private vehicles as large as the demand, the match rate from public vehicles seems negligible in the total match rate, compared to that from the private vehicles.

Moreover, the match rate for public vehicle actually decreases as private vehicle increases in supply. Even at the highe dispatch period of 1 minute, the match rate from public vehicle decreases from around 5 to 2%, from RSD of 10% to 100%, respectively. This means that as more private vehicles are available for rideshare, the passengers are taking less public vehicles and more private vehicles. A possible reason is the flexibility of private vehicle, which carry much more detour burden than public vehicles and satisfy the spatial and temporal constraints of passengers better. As well, the increase of match rate largely from RSD of 50 to 100% in general means that the match rate has not increased to a stable point. This shows us that public transportation does not affect the match rate around the critical mass of rideshare system, as will be discussed in more detail in Chapter 5.

The results from Figure 21 provide an insight in the implementation of rideshare that it may be helpful to mix the public transportation into the vehicle pool at a very low supply of private vehicles. However, when the private supply becomes as large as the demand at RSD of 100%, the benefits of public transportation becomes negligible because the dominant contributor of match rate becomes the private vehicles. Of course, the bus routes generated here are arbitrary and the results may vary from city to city and its transportation system. Therefore, a rideshare system must be evaluated on the integration of public transportation for real cities for practical application.

4.4 Conclusion

The use of public transportation of high reliability and low flexibility has been mentioned in a few rideshare studies for its potential [3], [16], [29]. In this chapter, the effect of public transportation to the vehicle pool is evaluated in terms of the match rate with the proposed framework. The results of this chapter shows different influence of public transportation on match rate at different levels of private vehicle supply. With very small supply of private vehicles at a low RSD, there is a slightly positive effect of increasing bus supply on the total match rate.

However, if the supply of private vehicle increases up to the size of demand, the effect of public transportation is more difficult to discern. One reason is that the increased match rate with private vehicle is proportionally much larger to the match rate with public vehicle so that the influence of public vehicle seems negligible in terms of the total match rate. Another reason is that as private vehicle is supplied more, more passengers take private vehicles and the match rate with public vehicle actually decreases.

Therefore, a rideshare system at a very low supply of vehicles may integrate public transportation as a complimentary vehicle pool for a slight boost of match rate in practice. However, that the effect of the public transportation must be evaluated before practical application as different settings of public transportation for different cities may change the results.

# Chapter 5. Schedule Flexibility of Rideshare

In this chapter, the effect of schedule flexibility on rideshare matching performance is evaluated. Schedule flexibility is the extra travel time a participant is willing to take to complete a journey, additional to the shortest travel time from the origin to destination. Described in detail in section 2.4, this relaxation feature loosens the temporal constraint of participants in forming a match. In addition to the other relaxation features used in building the matching framework, this temporal feature may be nudged into the rideshare service. For instance, to enlarge the schedule flexibility of participants, the input of time window of travelling may only be set at least some minutes larger than the direct travel time or a large time window can be given incentives to the participants. This parameter may affect the performance of matching efficiency and if so, its mechanism must be studied.

Therefore, it is necessary to evaluate how different distributions of schedule flexibility influence the matching efficiency so that a rideshare matching algorithm can maximize the use of a given set of requests. Following is the methodology and data used to evaluate schedule flexibility on the rideshare matching efficiency. Then, the results are analyzed to show that schedule flexibility influences positively on the matching efficiency. As well, increasing schedule flexibility is tested for a public transportation system to find the high potential of multi-modal rideshare system. This chapter is finished with a conclusion.

## 5.1 Methodology

To study the effect of schedule flexibility on rideshare matching, the match rate of a given request set is evaluated with increasing Ratio of Supply to Demand (RSD) for different distribution schedule flexibility. The terms used in this paper for evaluation are defined here. A critical mass, $RSD_{CM}$, is defined as the tipping point of participant size for performance stability, measured as the ratio of supply to demand where two conditions are both met. The two conditions are as follows.

- Condition 1. $RSD_{CM}$ is equal or larger than the first RSD value with Rate of match rate $\geq 0.50$ for a moving window with width of 100% RSD, starting from $RSD = 0\%$.

- Condition 2. $RSD_{CM}$ is equal or larger than the first RSD value with Rate of match rate $\leq 0.01$

for a moving window with width of 50% RSD, starting from RSD = 0%.

Where, rate of match rate = $\frac{\Delta match\ rate}{\Delta RSD}$, $unitless$. Also, critical match rate, $match\ rate_{CM}$, is defined as the match rate at the critical mass.

## 5.2 Data

The test network is same to the Chapter 3 with the simplified network of Sioux Falls shown in Figure 17. The rideshare matching parameters are 5 minutes for the transfer time, 10 minutes for the limit of walking, RSD of 10,50,100,150,200,250,300,350, and 400%. No public vehicle is used in this chapter to isolate the effects of schedule flexibility on match rate alone. The earliest departure time is randomly generated from 9AM to 10AM for all participants. The temporal flexibility for participants is randomly sampled from normal distribution with mean of 5,10,20,30,40, and 60 minutes and standard deviation of 10 minutes. The absolute values are used. Sample size is described for each figure of results. The seat capacity of each driver is 3 passengers and each passenger can take only one transfer, making two hops. For the Genetic Algorithm, the convergence threshold is 30 iterations, population size of 20, mutation rate of 0.01, and cross-over rate of 1 with single-point cross-over at random sampled location from uniform distribution along the chromosome span, fitness variable as the match rate, and the objective function to maximize the fitness variable.

## 5.3 Results

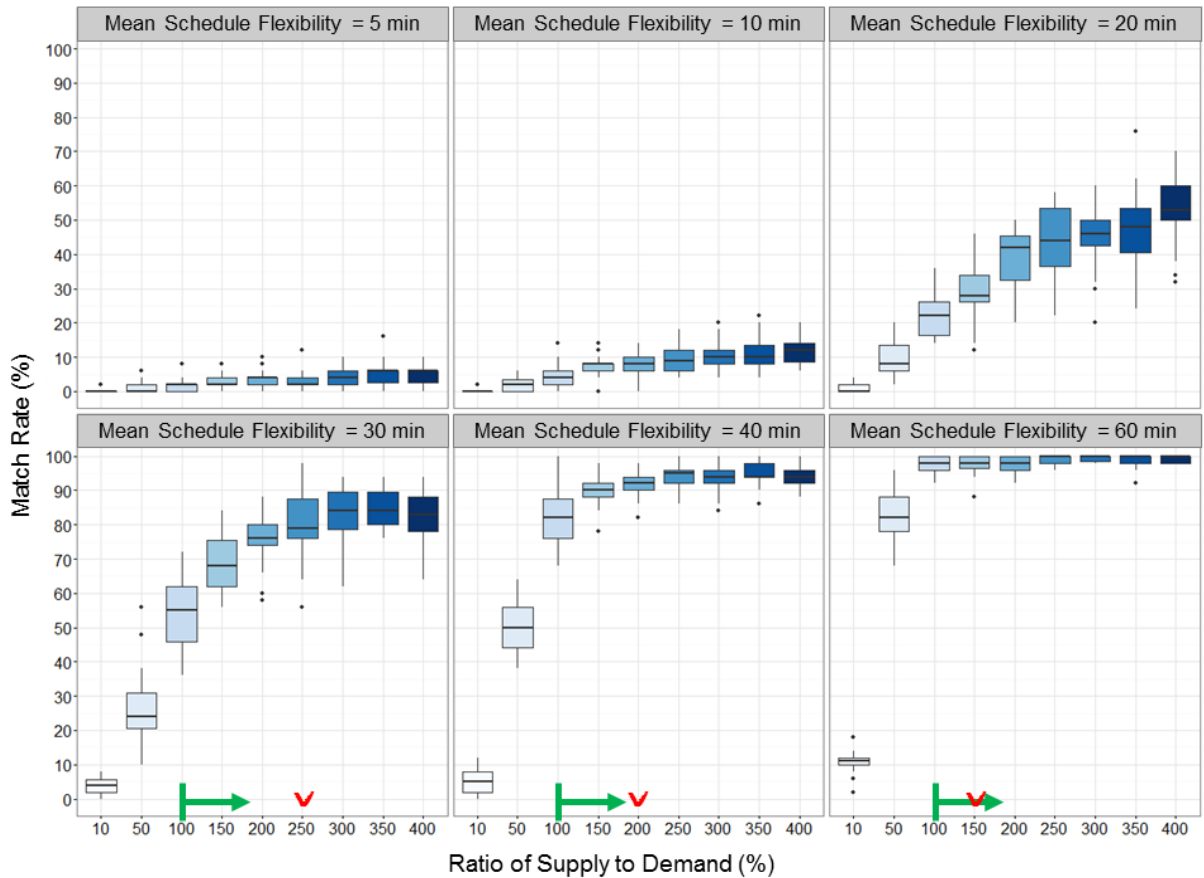

Figure 22 Match Rate for Different Schedule Flexibility

Table 8 Median Match Rate and Critical Mass for Different Schedule Flexibility

| Schedule Flexibility | RSD | Median Match Rate (%) | Rate of match rate | |
| --- | --- | --- | --- | --- |
| | | | Moving window width =RSD 50% | Moving window width =RSD 100% |
| **5 minutes** | 0 | 0 | - | - |
| | 10 | 0 | - | - |
| | 50 | 0 | 0 | - |
| | 100 | 2 | 0.04 | 0.02 |
| | 150 | 2 | 0 | 0.02 |
| | 200 | 4 | 0.04 | 0.02 |
| | 250 | 2 | -0.04 | 0 |
| | 300 | 4 | 0.04 | 0 |
| | 350 | 6 | 0.04 | 0.04 |
| | 400 | 6 | 0 | 0.02 |
| **10 minutes** | 0 | 0 | - | - |
| | 10 | 0 | - | - |
| | 50 | 2 | 0.04 | - |
| | 100 | 4 | 0.04 | 0.04 |
| | 150 | 8 | 0.08 | 0.06 |
| | 200 | 8 | 0 | 0.04 |
| | 250 | 9 | 0.02 | 0.01 |
| | 300 | 10 | 0.02 | 0.02 |
| | 350 | 10 | 0 | 0.01 |
| | 400 | 12 | 0.04 | 0.02 |
| | 0 | 0 | - | - |
| | 10 | 0 | - | - |
| | 50 | 8 | 0.16 | - |
| | 100 | 22 | 0.28 | 0.22 |
| 20 minutes | 150 | 28 | 0.12 | 0.2 |
| | 200 | 42 | 0.28 | 0.2 |
| | 250 | 44 | 0.04 | 0.16 |
| | 300 | 46 | 0.04 | 0.04 |
| | 350 | 48 | 0.04 | 0.04 |
| | 400 | 53 | 0.1 | 0.07 |
| | 0 | 0 | - | - |
| | 10 | 4 | - | - |
| | 50 | 24 | 0.48 | - |
| | 100 | 55 | 0.62 | **0.55** ① |
| 30 minutes | 150 | 68 | 0.26 | 0.44 |
| | 200 | 76 | 0.16 | 0.21 |
| | 250 | 79 | **0.06** ② | 0.11 |
| | 300 | 84 | 0.1 | 0.08 |
| | 350 | 84 | 0 | 0.05 |
| | 400 | 83 | -0.02 | -0.01 |
| | 0 | 0 | - | - |
| | 10 | 5 | - | - |
| | 50 | 50 | 1 | - |
| | 100 | 82 | 0.64 | **0.82** ① |
| 40 minutes | 150 | 90 | 0.16 | 0.4 |
| | 200 | 92 | **0.04** ② | 0.1 |
| | 250 | 95 | 0.06 | 0.05 |
| | 300 | 94 | -0.02 | 0.02 |
| | 350 | 94 | 0 | -0.01 |
| | 400 | 94 | 0 | 0 |
| | 0 | 0 | - | - |
| | 10 | 11 | - | - |
| | 50 | 82 | 1.64 | - |
| | 100 | 98 | 0.32 | **0.98** ① |
| 60 minutes | 150 | 98 | **0** ② | 0.16 |
| | 200 | 98 | 0 | 0 |
| | 250 | 100 | 0.04 | 0.02 |
| | 300 | 100 | 0 | 0.02 |
| | 350 | 100 | 0 | 0 |
| | 400 | 100 | 0 | 0 |

Figure 22 shows the boxplot of the match rate for different schedule flexibility with increasing Ratio of Supply to Demand at a given request set. There are six graphs of different schedule flexibility values, all with x-axis as RSD values in percentage, and y-axis as maximum match rate reached in each rideshare matching sample of GA. Note that the x-axis is not to scale. Table 8 lists the median values and the rate of match rate for each boxplot of Figure 22, to verify how critical mass is selected. The two conditions for critical mass are indicated in both the figure and table. Condition 1 is indicated by a green arrow and a green symbol ① and Condition 2 is indicated by a red tick and a red symbol ② in the figure and table, respectively. The sample size is 30 for each value of RSD and of schedule flexibility, making up total of 30 samples × 9 RSDs × 6 values of schedule flexibility = 1620 samples.

First observation from the figure is the increase of match rate with larger vehicle supply, which agrees with the result from section 4.3. A critical mass is reached at RSD below 400% for schedule flexibility of 30 minutes and longer shown with red ticks on graphs. It is possible that a critical mass can be achieved at larger RSD for schedule flexibility of less than 30 minutes. However, supply more than 4 times larger than the demand is already a very large pool of vehicles and may discourage more drivers to join the system because only a quarter of the drivers can be assigned even at the maximum.

Also, the match rate is larger at a larger schedule flexibility shows at any RSD values. However, for different values of schedule flexibility, the match rate increases at a different rate with increasing RSD. This means that the value of critical mass decreases with larger schedule flexibility, in other words, the match rate reaches a stability at a lower supply level. For instance, $RSD_{CM}$ decreases from 250%, 200%, and to 150% at mean schedule flexibility of 30, 40, and 60 minutes, respectively. The match rate increases more rapidly with increasing supply if the schedule flexibility is larger. Therefore, the system finds the critical mass faster at a smaller supply to demand ratio.

Third observation from the figure is that even as the supply at the critical mass falls with larger schedule flexibility, the critical match rate does not fall. Better yet, it increases. For instance, the $match\ rate_{CM}$ increases from 80, 90, and to 100% for mean schedule flexibility of 30, 40, and 60 minutes, respectively. This is important because with larger schedule flexibility, a smaller critical mass is found with only larger match rate, rather than having a trade-off relationship between the size of critical mass and the critical match rate.

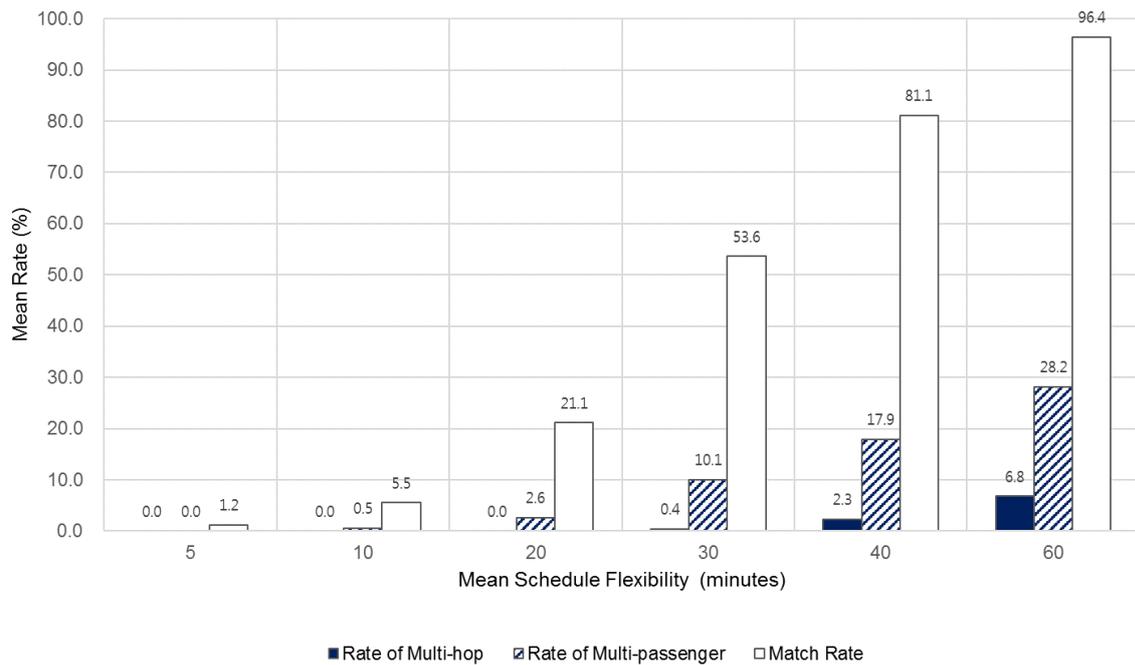

**Figure 23 The Effect of Schedule Flexibility on Multi-hop and Multi-passenger Rides**

To understand the mechanism of how schedule flexibility affects the match rate, Figure 23 is shown to describe percentages of multi-hop and multi-passenger rides for different schedule flexibility values. Each bar shows the mean rate values for different schedule flexibility. The bars with solid fill indicate the rate of multi-hop rides, calculated by dividing the number of passengers with multi-hop rides with total number of passenger request, n. The hatched bars indicate the rate of multi-passenger rides, calculated by dividing the number of passengers taken by drivers that serve more than one passenger, with total number of passenger request, n. The white bars indicate the match rate as equation (1). The number of passenger requests is 50 and RSD of 100%, with no public transportation. The total sample size is 30 samples × 6 schedule flexibility values = 180 samples.

The three different bar types represent the successful rides of multi-hop, multi-passenger, and single-hop and single-passenger. From the graph, the single-hop and single-passenger rides are represented the gap between the total match rate and sum of multi-hop and multi-passenger rates. This single-hop and single-passenger rides increases most rapidly, as schedule flexibility increases. For instance, the gap is about 19% but about 61% at schedule flexibility of 20 minutes and 40 minutes, respectively. The reason why single-hop and single-passenger rides increases the fastest is possibly because it is the simplest form of rideshare matching. An increase of schedule flexibility for both passenger and driver has affected this configuration the most.

Second, the increasing rate of multi-passenger rides is larger than that of multi-hop with larger schedule flexibility. For instance, the rate difference is around 10% for schedule flexibility of 30 minutes but increases to 21% for schedule flexibility of 60 minutes. This is possibly due to the subject of temporal scheduling. As explained in 3.3.2.1, the temporal scheduling is planned in driver's perspective rather than the passenger's, where the time schedule of all visits are planned for each driver and his course of driving. That is why larger schedule flexibility affects more on the multi-passenger rides, which are match expansion of driver's perspective of ridesharing, than the multi-hop rides, which are match expansion of passenger's perspective of ridesharing. Because the time schedule is planned independently between drivers, the chance of multi-hop may be difficult to increase with the relaxation of temporal constraint.

Therefore, it is possible to maximize the use of schedule flexibility if it is larger for drivers than the passenger among the participants. This means that if the passengers have larger schedule flexibility than the drivers, the temporal scheduling may bring less efficient results with driver-oriented temporal scheduling. For a practical application, a rideshare system must investigate which role of participants is more flexible in terms of the time schedule to give an insight whether the temporal scheduling must be done in terms of the driver or passenger.

Additionally, it is possible to compare the performance of rideshare system with the public transportation system for different schedule flexibility of riders. The former system may use private vehicles only and the latter may use the public vehicles only, to see which system satisfies the passenger's trip constraints better. The core difference between the two systems is the detour burden on passenger. Though the public transportation data used in chapter 4 is generated data for the simplified network, it is possible to assume a high flow of bus to compare with the private vehicles.

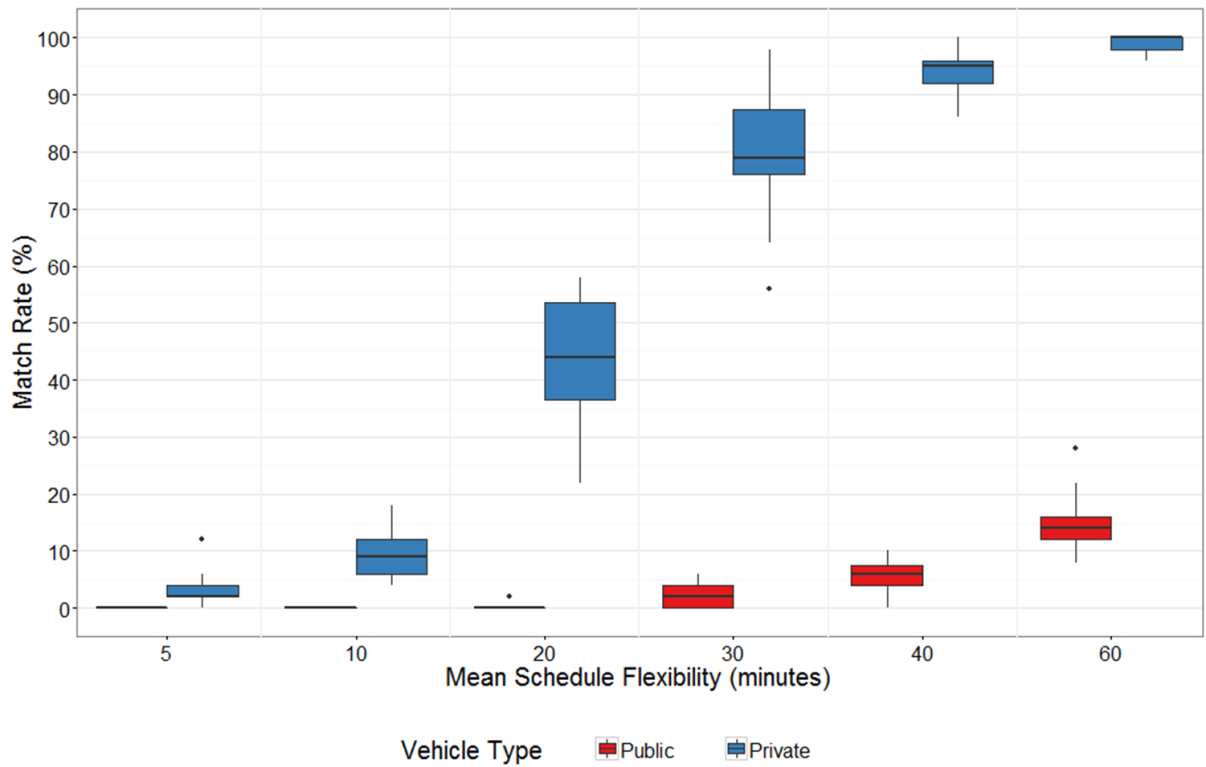

**Figure 24 Match Rate of Public Transportation for Different Schedule Flexibility**

The results shown in Figure 24 compares the successful rides of public transportation with that of rideshare with private vehicles, in terms of match rate. The private vehicle in blue is the reproduced boxplots of RSD as 250%, same to those in Figure 22. The public transportation in red use a total of 9 bi-directional bus routes with fixed dispatch period of 5 minutes for the simplified network of Sioux falls, similar to Chapter 4. The 5-minute dispatch period represents flow of 143 buses per minute in the system, which is 286% of the passenger requests and larger than the private vehicles. Note that this is flow per minute, so the request environment actually has much more vehicles available to use per matching algorithm execution.

Comparing to the rideshare with private vehicles in blue, the public vehicles satisfy rides of much less passengers, even at a high frequency of dispatch. At a low schedule flexibility of 5 to 10 minutes, the difference does not seem great. As schedule flexibility increases, however, the difference between rideshare of private vehicles and usage of public transportation increases rapidly. When mean schedule flexibility reaches as large as 60 minutes, the rideshare with private vehicles reaches matching efficiency of almost 100%, however the public transportation lingers well below 20%.

The low success of public transportation is due to the rigid routes of the public vehicles that do not take any detour burden, which is one of the critical disadvantages of public transportation. Therefore, the flexible routes of rideshare has much higher match rate. This shows that the proposed framework of multi-modal rideshare may be used to increase the flexibility of conventional public transportation system. In addition, this framework can be used to evaluate the performance of public transportation design with various objective functions, for instance with satisfied trips represented as the match rate shown in Figure 24.

5.4  Conclusion

This chapter evaluates the effect of schedule flexibility on matching efficiency rate using the rideshare matching framework. Schedule flexibility is the extra time the participants are willing to travel additional to the direct travel time by car and reduces the temporal constraint of rideshare matching. The match rate of increasing Ratio of Supply to Demand is evaluated for different distribution of schedule flexibility. Public vehicle is not used in this study to eliminate the effect of vehicle pool in the evaluation.

The results in Figure 22 show that with larger schedule flexibility, the match rate increases much more rapidly with the RSD. This effect results in the reduction of critical mass, $RSD_{CM}$, with larger flexibility. Also, it is shown that the reduced critical mass has better rideshare performance, indicated by larger critical match rate, $match\ rate_{CM}$. This shows that the stability of match rate after a smaller critical mass is not in a trade-off relationship with the critical match rate. Therefore, schedule flexibility can be nudged into the system with incentives or request requirement to boost the performance of the rideshare matching.

Additionally, the mechanism of how schedule flexibility affects match rate is further studied by evaluating the increase of multi-hop and multi-passenger matches separately. In Figure 23, it is shown that the simplest form of rideshare match, single-hop and single-passenger, increases the most with larger schedule flexibility of both passenger and driver. Another result is that the subject of temporal schedule is affected more with the schedule flexibility. In this framework, the temporal schedules are planned in driver's perspective and this design explains the larger increase of multi-passenger rides than the multi-hop rides, since multi-passenger rides are the expansion of driver's ridesharing constraints. This result gives an insight to a rideshare system designer that the temporal planning of rideshare must pay attention to which role of the rideshare participants has more schedule flexibility in order to maximize its positive effect on the rideshare matching efficiency.

In addition, a public transportation system is compared to the rideshare system with private vehicles in terms of the number of satisfied trips with increasing schedule flexibility. The result in Figure 24 shows that at a given set of travel demand, the public transportation system facilitates a much lower number of passenger trips than the rideshare system with private vehicles. The discrepancy between the performance of public system and private vehicle system for mobility becomes even larger as the schedule flexibility of passengers increases. This is due to the rigid routes of public transportation that takes no detour burden and shows for a need to develop a flexible transportation system, which can be facilitated by the fusion of private and public vehicles for mobility as developed in this research.

# Chapter 6. Conclusion

In this study, a multi-modal matching framework is developed that opens up a new dimension of shared mobility with multiple mode choices, evolved from a conventional rideshare system. Rideshare is a joined trip of drivers and riders, whose requests are matched to meet every participant's need of travel. This system has benefits of increasing efficiency of vehicles and mobility of travelers without vehicles [1]–[32]. With GPS-equipped mobile phones, a rideshare system has a better advantage of using accurate location information of participants in real time.

However, one of the challenges faced in its implementation is the chicken-and-egg problem, where drivers and riders are co-dependent in terms of the matching efficiency [16], [22], [23], [28]. It has encouraged the researchers to develop a rideshare matching algorithm that has a successful matching efficiency even with a low ratio of supply to demand of drivers. As studied in Chapter 2, however, the current literature lacks a study on a matching algorithm of multi-modal trips from allowing multi-hop rides and additional vehicle pool. This framework can largely expand the conventional rideshare system to a mobility share system with different mode choices and journey types and maximize the matching efficiency even further.

Therefore, Chapter 3 develops a multi-modal matching framework, which fully relaxes match configuration, detour burden, and vehicle pool augmentation of rideshare rule. The framework allows multi-modal rides with public transportation, whose detour is realistically designed for different vehicle type of drivers. This framework successfully finds match partners that maximize the match rate using Genetic Algorithm.

Also, the effect of vehicle pool augmentation with public transportation is thoroughly studied in Chapter 4 to provide an insight in the practical implementation of rideshare. Vehicle pool augmentation with public transportation has only been mentioned of its potential but never studied on its impact [3], [16], [29]. The results show that public transportation may improve the match rate slightly at a very low supply of private vehicle. The benefit of public vehicles to the rideshare pool becomes negligible compared to the private vehicle as more private vehicles join. However, the effect of public vehicle on rideshare must be evaluated for different cities as it may affect the results.

As well, the schedule flexibility of participants is studied in terms of the matching efficiency and critical mass in Chapter 5. In the results, it is evident that the match rate increases with increasing supply but more rapidly when the schedule flexibility of participants is higher. Therefore, the critical mass at which the match rate becomes stable is smaller with larger schedule flexibility. Also the match rate at the reduced critical mass increases, so that the matching performance is not degraded with more stability.

Digging deeper on how the schedule flexibility can be fully utilized in terms of the match rate, it is found that the subject of temporal scheduling is affected more with the temporal relaxation of larger schedule flexibility. Multiple-passenger rides are the expansion of rideshare constraint of drivers, at the same time the time schedule is planned in terms of the drivers. The results describe that multiple-passenger rides are increased more than multi-hop rides with larger schedule flexibility. Therefore, a rideshare matching system must pay attention to which role of rideshare has more schedule flexibility in the requests to maximize its benefits on the matching efficiency.

The contribution of this research is further developed by comparing the proposed rideshare system with a public transportation system. It is shown that fulfilment of mobility demand is much higher for the rideshare system with private vehicles than the public transportation system, even though the number of public vehicles available is larger than the private vehicles. This effect is contributed to the rigidity of public vehicle routes, where no detour burden is accepted. Therefore, this result confirms the need to develop a flexible transportation system, which can be facilitated by the fusion of private and public vehicles for mobility as developed in this research.

For future work, it is possible to test the expandability of this framework. This framework can be easily applied for different shared-mobility systems, such as shared autonomous vehicles and dynamic demand-responsive public vehicles, as its constraints of drivers can be easily relaxed. Also, the framework can be evaluated with other objective functions, such as minimizing the total vehicle mile travelled, maximizing utility of participants, or multi-objective functions [3], [4], [6]–[8], [12], [15], [18]–[20], [26], [37]. The objective function used in the framework is the maximization of match rate in the operational perspective, however the framework can easily adopt to other objective functions in the participants' perspective by simply switching the objective function of GA. Also the framework can further be improved by optimizing the route of the joined trip within the framework, which can reduce the travel time and increase the chance of forming a successful match. To study the feasibility of implementing this framework to real world, real data can be applied to find rideshare matches and test its computing time solved with distributed computing. A strategy of dividing its computing load can be devised, such as section-based calculation where driver candidates are filtered for each passenger by its section of the city. As well, speed prediction could be considered to calculate a more accurate travel time, thereby reducing the uncertainty of visit times of rideshare.

# Acknowledgement

3년 전 진로상담을 하러 교수님을 찾아 뵌 적이 기억납니다. 처음 보는 학부생인데도 불구하고 한 시간 동안 성심 성의껏 상담을 해주신 우리 교수님. 그 모습에 감동을 받고 연구실에 들어와보니 선배들 또한 마음씨 좋고 배울 것이 많다는 것을 알게 되었어요. 훌륭한 학문 지도와 따뜻한 인성 지도를 받고 싶어서 저는 석사 유학 준비를 접어 연구실에 들어왔고, 벌써 논문의 맨 뒤 감사의 글을 쓰고 있습니다.

우리 연구실에 있는 동안 정말 많은 것을 배웠습니다. 3년 전에는 코딩도 몰랐고, 논문의 목적도 몰랐고, 또 같이 일을 할 때 갖춰야할 많은 것들에 대해 무지했습니다. 지금도 많이 부족하고 배워야할 게 태산이지만 석사 졸업이라는 잠깐의 뒤돌아봄이 있어 여태까지 배운 것에 감사하는 마음을 전하게 될 수 있게 되었네요.

교수님! 평생 스승님 이라고 부를 저의 첫 번째 지도 교수님. 저를 믿어주시고, 격려해주시고, 도전하게 해주셔서 정말 감사합니다. 실수도 하고 부족한 점이 많은 제가 교수님께 받은 것은 큰 사랑과 참된 교육이라고 생각합니다. 제가 이렇게 성장하도록 처음부터 저를 어떻게 그리 믿어 주셨는지 모르겠어요. 어려운 상황에서도 밝게 생각하고 더 열심히 하고자 하는 마음을 심어 주시는 교수님 덕분에 더 큰 도전을 할 의지와 목표가 생겼습니다. 3년 동안 제가 배우고 이룬 모든 것에 대해 감사합니다. 앞으로 어딜 가서도 굴하지 않고 저의 의지와 가치와 열정을 갖고 옳은 선택하는 제자가 되겠습니다. 평생 갚아야할 은혜를 잊지 않을게요.

그리고 우리 연구실 분들! 하나하나 얼굴을 떠올려 보았더니 얼굴에 미소가 떠오릅니다. 먼저 세현오빠. 몇 년 전 디자인 수업 때 조교인 오빠를 귀찮게 했을 때부터 알아봤어야 했나 봐요. 이렇게 많은 일을 같이 하고, 많은 논문을 쓰게 될지 줄 누가 알았을까요. 어렵고 부족한 일이 있으면 언제나 도와주고 채워준 오빠, 덕분에 졸업 전에 좋은 논문도 쓰고 상금도 타고 디펜스도 잘 끝낼 수 있었어요. 오빠 정말 감사드리고, 평생 좋은 이모로 아연이 가족에게 갚을게요! 그리고 동훈오빠, 종해오빠. 꼼꼼하게 봐준 똑똑한 오빠들 덕분에 석사 논문 컨트리뷰션도 좋아지고 결과도 훨씬 보기 좋아진 것 같아요. 바쁜 와중에도 성심성의껏 도움 주신 마음 정말 감사해요. 동훈오빠는 운동으로 튼튼맨이 되고, 종해오빠는 칼퇴근 하는 날이 어서 왔으면 좋겠네요. 그리고 성훈오빠. 같이 일하느라 고생 많으셨어요. 묵묵히 연구실을 위해 일하는 오빠 덕분에 석사생활이 훨씬 좋았었다고 믿어요, 감사합니다. 자기 일에 최선과 책임을 다하는 모습 진짜 멋있어요! 성준오빠. 언제나 침착하고 신중한 오빠 모습은 제가 정말 배워야 할 것 같아요. 어려운 일이 있으면 누구보다 먼저 발 벗고 도와주는 오빠, 언제나 감사해요. 시몬오빠. 특유의 끈질김과 집중력으로 포닥 잘 끝내고 오시리라 믿어요. 오빠가 없으니 연구실이 조용하네요. 어서 한국으로 돌아와 휴스턴에게 인사하러 갑시다.
62

용준오빠. 어려운 문제에는 그 무엇보다 명쾌한 혜안을 갖고 있는 용준옹! 언제나 오빠처럼 솔루션을 갖고 있는 솔로몬이 되고 싶어요. 민주오빠. 연구실을 나가서도 챙겨주시는 오빠, 그 마음 정말 감사드려요. 예쁜 나윤이도 무럭무럭 자라길 바랄게요. 수빈언니. 연구실 처음 들어왔을 때 단둘이 여자라 의지가 많이 되었어요. 회사 바쁘게 다니면서도 꼭 챙겨주고 열심히 하라며 응원해주는 언니, 너무 감사해요. 진현이. 자기 일에 집중하고 똑똑하게 해결하는 모습 정말 멋있어. 열심히 하는 모습 언제나 기대할게. 은혜언니. 짧게나마 내 짝꿍이 되어서 나에게 얼마나 힘이 됐는지 몰라. 내 디펜스 열심히 하라고 하느님이 은혜를 베푸셨나봐. 미국 가서 언니 열심히 하는 모습 보러 꼭 갈게. 그리고 준용, 수형, 정윤이. 처음 같이 연구실 들어와서 서로 의지가 많이 되고 배운 것도 많아, 고마워. 혼자 있었다면 외로워 했을 텐데 같이 있어서 좋은 추억을 만들 수 있었어. 지금은 다 각자 갈 길을 가고 있지만, 맡은 일에 충실하고 좋은 일만 생기길 바랄게. 성진이. 우리 연구실을 강력히 추천 했었는데 역시 들어와서 잘 적응을 하는 것 보기 좋다. 선후배랑 같이 열심히 노력해서 어서 좋은 성적 낼 수 있었으면 좋겠어, 기대할게. 예은이. 우리 예은이도 꼼꼼한 성격으로 재미있는 연구를 많이 선사해줬으면 좋겠어. 꾸준히 열심히 해서 좋은 결과 볼 수 있다고 믿어.

또 힘들 때나 좋을 때나 꼭 생각나는 한영이랑 성헌이. 언제나 나의 편이 되어주는 든든한 수민이와 희정이. 나와 마음이 제일 잘 맞는 현경언니. 그리고 내게 언제나 긍정 에너지를 선사하는 지원이. 모두 나에게 큰 힘이 되어주었고, 정말 감사드려요.

끝으로 나를 언제나 믿고 사랑해주는, 내 자신보다 더 소중한 나의 가족에게 감사드립니다. 먼저 나의 꿈을 키워주고 지켜준 아빠와 엄마, 막내딸까지 뒷바라지하느라 고생시키는 것 같아 미안하고 감사해요. 더 많은 도전이 기다리지만 아빠랑 엄마가 나에게 보여준 정직하고 성실한 모습만 닮아 노력할게요. 그리고 나를 위해 희생을 많이 했을 지민언니와 승민언니, 언제나 응원하고 기도해주시는 외할머니와 친할머니, 말이 필요 없는 나의 분신 이수린. 그리고 내 남동생 우해롱까지. 앞으로 나의 모든 것을 다 바쳐 가족에게 봉사하도록 할게요. 사랑합니다.

우수민 올림.

# Curriculum Vitae

## EDUCATION

Korea Advanced Institute of Science and Technology (Daejeon, Korea)         2009 Fall – 14 Spring

- Undergraduate
- Major: Civil and Environmental Engineering
- Minor: Science Technology and Policy
    - Overall Grade: 4.02/4.3 (Summa cum laude)

Korea Advanced Institute of Science and Technology (Daejeon, Korea)         2014 Fall – 16 Fall

- Master's Degree
- Smart Transportation Systems Laboratory
- Department of Civil and Environmental Engineering

## RESEARCHES/STUDIES

### Papers and Conferences

- Sampling-based Collision Warning System with Smartphone in Cloud Computing Environment - 2015 June
    - 2nd Author
    - The IEEE Intelligent Vehicles Symposium 2015

- Optimization of Pavement Inspection Schedule with Traffic Demand Prediction - 2015 August
    - 1st Author
    - 11th International Conference of the International Institute for Infrastructure Resilience and Reconstruction (Conference)
    - Procedia – Social and Behavioral Sciences (Journal)

- Day-based Prediction of Origin-Destination Demand of Large Network - 2015 November
    - 1st Author
    - KKHTCNN Symposium on Civil Engineering

- Data-driven Imputation Method for Traffic Data in Sectional Units of Road Links - Accepted in 2015 November and in press
    - 2nd author
    - IEEE Transactions on Intelligent Transportation Systems

- Development of data-driven prediction methodology of origin-destination demand in large network for real-time services - 2016 January
    - 1st author
    - Transportation Research Board 2016 Annual Meeting (Conference)
    - Journal of Transportation Research Record (Journal)

### Awards

- Presidential Award by Korean Society of Transportation (2014 February)
- Certification of Commendation by Korean Society of Civil Engineers (2014 February)
- Best Presentation Award at 28th KKHTCNN Symposium on Civil Engineering (2015 November)
- Second Place at U-City Contest of Research (2016 January)

### Projects

- Microsoft Azure for Researchers (2014)
  - Presented at Microsoft Research Asia: Faculty Summit of 2014, Beijng

- Software Development Weather-based Big Data Service (2014 Fall – 2015 Spring)
  - By Korea Meteorological Administration

- Microsoft Azure for Researchers (2015)

### Patent and Copyrights

- Patent: Apparatus and Method for Correcting Transportation Data (December 28$^{th}$, 2015)
- Software Copyright: Pattern-matching based Origin-Destination Traffic Demand Prediction Program)
- Software Copyright: Online Simulation-based Optimal Road Control Program (January 5$^{th}$, 2016)
- Software Copyright: Human Behavior-based Train Simulation Program (January 5$^{th}$, 2016)

## REFERRENCE

- Associate Professor. Hwasoo Yeo (Department of Civil and Environmental Engineering, Korea Advanced Institute of Science and Technology)